\documentclass[pre,showpacs,twocolumn,superscriptaddress,floatfix,amsmath]{revtex4}
 \usepackage[latin1]{inputenc}
 \usepackage[english]{babel}
 \usepackage{graphicx}
 \usepackage{psfrag}
 \usepackage{amssymb}
 \usepackage{amsmath}
 \usepackage{amscd}
 \usepackage{eucal}
 \usepackage{color}
 \usepackage{bm}
 
 \newcommand{\ignore}[1]{\relax}

\newcommand{\de}{\delta}
\newcommand{\De}{\Delta}
\newcommand{\eps}{\epsilon}

\newcommand{\tEc}{\tau_{\rm E}^{\rm c}}
\newcommand{\tEo}{\tau_{\rm E}^{\rm o}}

\newcommand{\tD}{\tau_{\rm D}}

 \newcommand{\e}{{\rm e}}
 \newcommand{\rmd}{{\rm d}}
 \newcommand{\rmi}{{\rm i}}
 
 \newcommand{\tr}{{\rm tr}}

\begin{document}

\title
{Semiclassical transport in nearly symmetric
quantum dots II: \\
symmetry-breaking due to asymmetric leads}

\author{Robert S. Whitney}
\affiliation{Institut Laue-Langevin, 6 rue Jules Horowitz, B.P. 156,
         38042 Grenoble, France}

\author{Henning Schomerus}
\author{Marten Kopp }
\affiliation{Department of Physics, Lancaster University, Lancaster, LA1 4YB, United Kingdom}

\date{June 4, 2009}

\begin{abstract}
In this work --- the second of a pair of articles --- we consider 
transport through spatially symmetric quantum dots with leads whose widths
or positions do not obey the spatial symmetry.
We use the semiclassical theory of transport to find the 
symmetry-induced contributions to
weak localization corrections and universal conductance
fluctuations for dots with left-right, up-down, inversion and four-fold symmetries.
We show that all these contributions are suppressed by asymmetric leads,
however they remain finite whenever leads intersect with their
images under the symmetry operation. For an up-down symmetric
dot, this means that the contributions can be finite even if one of the leads is
completely asymmetric.
We find that the suppression of the contributions to universal conductance fluctuations is the square of the suppression of contributions to weak localization.  
Finally, we develop a random-matrix theory model
which enables us to numerically confirm these results.
\end{abstract}

% 05.45.Mt Quantum chaos; semiclassical methods
% 74.40.+k Fluctuations (noise, chaos, nonequilibrium superconductivity,
% localization, etc.)
% 73.23.-b Electronic transport in mesoscopic systems
% 03.65.Yz Decoherence; open systems; quantum statistical methods

\maketitle

%%%%%%%%%%%%%%%%%%%%%%%%%%%%%%%%%%%%%%%%%%%%%%%%%%%%%%%%%%%%%%%%
%%%%%%%%%%%%%%%%%%%%%%%%%%%%%%%%%%%%%%%%%%%%%%%%%%%%%%%%%%%%%%%%

\section{Introduction}

This is work --- the second of a pair of articles ---
on mesoscopic transport
through chaotic quantum dots with spatial symmetries (see Ref.
\cite{WSK1} for part I). In both works we use recent advances in
semiclassical techniques to address the effect of spatial
symmetries on weak localization (WL) corrections and universal
conductance fluctuations (UCFs). The aim of the first article was to
identify the microscopic origin of properties that were earlier
only known from phenomenological random-matrix theory (RMT)
\cite{Baranger-Mello,Gopar96,Gopar-Rotter-Schomerus,Kopp-Schomerus-Rotter},
and furthermore to extend the considerations to situations in
which RMT is not easily applicable. In particular, this includes
scenarios where symmetries are only partially preserved.
To this end, the first article \cite{WSK1} also considered the combined
effects of magnetic fields, a finite Ehrenfest time, and dephasing
on symmetric systems and also discussed the reduction of
symmetry-related interference effects by deformations of the dots.

\begin{figure}
\centerline{\hbox{
\includegraphics[width=0.73\columnwidth]{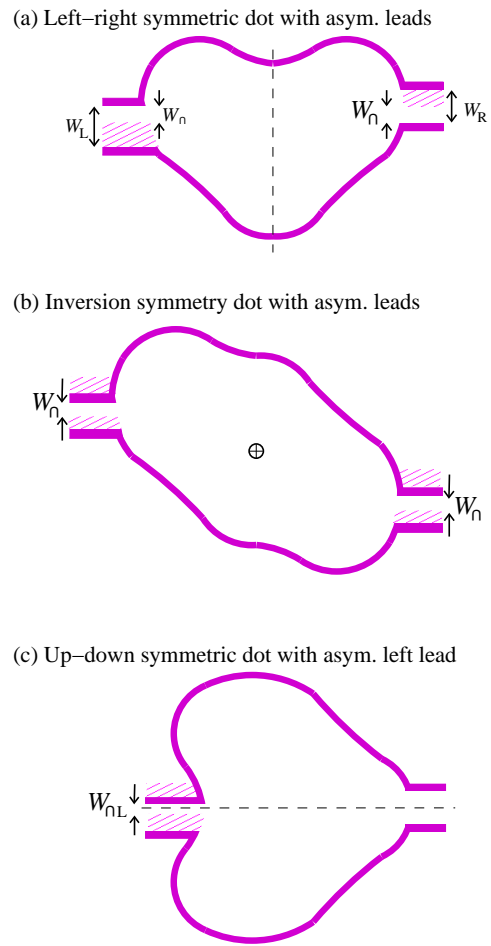}}}
\caption[]{\label{Fig:asymmleads} 
(colour online). (a) A quantum dot with a
left-right mirror symmetry, coupled to leads which do not respect
that symmetry. The left lead (L) has width $W_{\rm L}$, the right
lead (R) has width $W_{\rm R}$. The intersection between lead L
and the mirror image of lead R has width $W_\cap$. If the L and R
leads have no intersection under the mirror-symmetry then
$W_\cap=0$. (b) Same for a quantum dot with inversion symmetry.
(c) A quantum dot with up-down symmetry, for which each
symmetry-respecting lead is mapped onto itself. In the figure, the
left lead is displaced, which reduces the intersection $W_{\cap
L}$ of this lead with its mirror image.}
\end{figure}

In the present paper, we contrast this `internal' symmetry
breaking with symmetry breaking which is due to the position or
shape of the leads (for  examples of such  situations see
Fig.~\ref{Fig:asymmleads}). We ask what happens to the transport
 if we take a symmetric dot coupled to leads which
respect the symmetry, and then start moving one of the leads. In
the fully symmetric situation, the magnitude of UCFs is doubled
for each independent symmetry, while the weak localization
correction can be either increased or reduced (sometimes remain
unaffected) depending on the spatial symmetry in question
\cite{WSK1,Baranger-Mello,Gopar96}. Are these symmetry-induced
effects modified when the leads are deformed or displaced? If so,
are they sensitive to displacement on a quantum scale (of order of
a Fermi wavelength) or a classical scale (of order of a lead
width)?

%%%%%%%%%%%%%%%%%%%%%%%

The present literature does not offer much guidance to answer
these questions---indeed, the knowledge on transport in spatially
symmetric systems with displaced leads is rather limited.
Reference~\cite{Martinez-Mello} reports that the distribution of
transmission eigenvalues of a left-right symmetric dot with
completely asymmetrically-placed leads differs slightly from the
distribution of a completely asymmetric dot. Because the
difference is small, symmetric systems (such as stadium billiards)
with displaced leads are indeed often used as representatives of
completely asymmetric systems (see, e.g., Refs.\
\cite{Gopar-Rotter-Schomerus,Kopp-Schomerus-Rotter}).
 Recent works of one of the authors, on the
other hand, identify a huge conductance peak in weakly coupled
mirror-symmetric double-dots which still remains large even when
the leads are not placed symmetrically \cite{WMM,WMM2}.

A simple consideration of weak localization quickly convinces us
that it could never be as robust as the above-mentioned huge
conductance peak in double dots. In systems without spatial
symmetries, weak localization is the counter-part of coherent
backscattering---particle conservation guarantees that one cannot
have one without the other. Systems with spatial symmetries have
addition coherent back- and forward-scattering contributions (as
discussed in the first of this pair of articles \cite{WSK1}).
These contributions rely on interference between paths that are
related by  spatial symmetry. If those paths do not both couple to
the leads, they cannot generate an interference contribution to
conductance. Thus, if we displace one lead so much that there is
no intersection with its spatially symmetric partner ($W_\cap=0$
in Fig.~\ref{Fig:asymmleads}) then the contributions to coherent
forward scattering due to the spatial symmetries must vanish.

The precise distance by which one has to move the lead to
substantially suppress the symmetry-related contributions depends
on the detailed position dependence of the coherent forward- and
backscattering peaks. In principle, these coherent interference
patterns could oscillate on a scale of a wavelength, and thus one
might imagine that a small displacement of that order would
suffice. The calculations and numerical computations presented by
us here show that this is not the case. Instead, the coherent
forward- and backscattering peaks have a width of order the lead
width, and do not oscillate on the scale of a wavelength.

These considerations entail that the displacement of leads in
internally symmetric systems offers a unique means to study
coherent forward- and backscattering processes. From photonic
systems it is known that the shape of the coherent backscattering
cone provides valuable information on the multiple scattering in a
sample \cite{backscattering1,backscattering2}. Based on the
results of the present work, transport measurements with gradually
displaced leads promise to give similar insight into the dynamics
of electronic systems.

This work is organized as follows. Section \ref{sec:backgrd}
introduces notation and provides a condensed review of the basic
semiclassical concepts elaborated in more detail in the first of
this pair of articles \cite{WSK1}. The following sections describe
the consequences of displaced leads for the weak localization
correction in systems with left-right symmetry (Sec.\
\ref{sect:LR}), inversion symmetry (Sec.\ \ref{sect:inv}), up-down
symmetry  (Sec.\ \ref{sect:UD}) and four-fold symmetry (Sec.\
\ref{sect:4F}). In Sec.~\ref{sect:UCFs} we study the magnitude of
universal conductance fluctuations for all types of symmetry.
Finally, in Section~\ref{sect:RMT} we generalize the
phenomenological RMT model of symmetry breaking  (presented in
\cite{WSK1}) to the case of displaced leads, and compare the
results of numerical computations to the semiclassical
predictions. Our conclusions are collected in
Section~\ref{sect:conclusions}. The appendix contains some further
details on the semiclassical calculation of universal conductance
fluctuations.

\section{\label{sec:backgrd}Background}

%%%%%%%%%%%%%%%%%%%%%%%%%%%%%%%%%%%%%%%%%%%%%%%%%%%%%%%%%%%%%%%%%%%%%%%%%%%%%
%%%%%%%%%%%%%%%%%%%%%%%%%%%%%%%%%%%%%%%%%%%%%%%%%%%%%%%%%%%%%%%%%%%%%%%%%%%%%

To make this article self-contained we here first fix notation and
then briefly summarize the main concepts of the theory of
semiclassical transport in systems with spatial symmetries,
developed in the first of this pair of articles \cite{WSK1}.

\subsection{Characteristic scales}

We consider chaotic quantum dots of size $L$ [area $A={\cal
O}(L^2)$ and circumference $C={\cal O}(L)$] which may possess any
of the following three types of spatial symmetry; a left-right
mirror-symmetry, an inversion symmetry, and an up-down
mirror-symmetry. We also consider  four-fold symmetric systems
which simultaneously possess all the above symmetries. The quantum
dot is perfectly coupled to two leads, labelled left (L) and right
(R) and carrying $N_{\rm L}$ and $N_{\rm R}$ modes, where
$N_\kappa = p_{\rm F}W_\kappa/(\pi\hbar) \gg 1$ for $\kappa \in
{\rm L,R}$ (here $p_F$ is the Fermi momentum; we also denote the
Fermi velocity by $v_F$). The quantum dynamics in the dot is
characterized by a number of time scales, given by the time of
flight $\tau_0=\pi A/C v_F$ between successive reflections off the
boundaries, the dwell time $\tD = \tau_0 \times C/(W_{\rm
L}+W_{\rm R})$, the dephasing time
 $\tau_\phi=1/\gamma_\phi$ (where $\gamma_\phi$ is
the dephasing rate), and a time scale $\tau_B=(B_0/B)^2\tau_0$ on
which a magnetic field destroys time-reversal symmetry. Here, $B_0
\sim h/(eA)$ is a characteristic field strength at which about one
flux quantum penetrates the quantum dot. In transport, the effect
of a magnetic field is felt at a smaller magnetic field
\begin{equation}
B_{\rm c}= a B_0\sqrt{\tau_0/2\tD}
% =a B_0\sqrt{(N_{\rm L}+N_{\rm R})/2M}
\label{eq:bc},
\end{equation}
where $a$ is a system-specific parameter  of order one
\cite{Beenakker-1997}. Furthermore, the quantum-to-classical
crossover is characterized by the open-system Ehrenfest time
$\tEo=\Lambda^{-1} \ln [W^2/(L \lambda_{\rm F})]$ and the
closed-system Ehrenfest time $\tEc=\Lambda^{-1} \ln [
L/\lambda_{\rm F}]$, where $\Lambda$ is the classical Lyapunov
exponent and $\lambda_{\rm F}$ is the Fermi wavelength
\cite{Schomerus-Jacquod}.

In contrast to Ref.\ \cite{WSK1}, we here consider the possibility
that the leads do not respect the symmetry of the dot. As shown in
Fig.~\ref{Fig:asymmleads}, the displacement from the
symmetry-respecting position is characterized by the overlap of
leads under the relevant symmetry operation. For left-right mirror
symmetry and inversion symmetry, this is the width  $W_\cap$  of
the intersection of a lead with the  image of the other lead. An
up-down symmetry maps each symmetry-respecting lead onto itself.
The displacement of lead L (R) is then characterized by the width
$W_{\cap L}$ ($W_{\cap R}$) of the intersection of this lead with
its own mirror image. In a four-fold symmetric system, the
displacement is characterized by the various widths of
intersections with respect to the individual symmetries
($W_{\rm \cap LR}$ for left-right mirror symmetry, $W_{\rm \cap inv}$ 
for inversion symmetry, $W_{\rm \cap UD:L}$ for up-down mirror
symmetry of lead L and $W_{\rm \cap UD:R}$ for up-down mirror
symmetry of lead R).

\subsection{Semiclassical theory of transport}

The semiclassical theory of transport \cite{Bar91,Bar93} expresses
the transport through a quantum dot in terms of classical paths
$\gamma,\gamma'$ which connect point $y_0$ lead L to point $y$ on
lead R. Summing over lead modes as in Ref.~\cite{Jac06}, the
dimensionless conductance (conductance in units of $2e^2/h$) is
given by
\begin{eqnarray}\label{Eq:g-doublesum}
 g &=& {1 \over 2 \pi \hbar}
  \int_{\rm L}\!\!{\rm d} y_{0}   \int_{\rm R} \!{\rm d} y
    \sum_{ \gamma, \gamma'}
     A_{\gamma} A_{\gamma'} \;\;
    e^{ i (S_\gamma- S_{\gamma'})/\hbar},
 \end{eqnarray}
where $S_\gamma= \int_\gamma {\bf p} \rmd {\bf r}$ denotes the
classical action of a path, and the amplitude $A_\gamma$  is
related to the square-root of the path's stability.

For most pairs of $\gamma$ and $\gamma'$ the exponential in
Eq.~(\ref{Eq:g-doublesum}) oscillates wildly as one changes the
energy or the dot-shape. Thus they make no contribution to the
average conductance (where one averages over energy, dot-shape, or
both). The contributions that survive averaging are those where
the pairs of paths have similar actions  $S_\gamma \simeq
S_{\gamma'}$ for a broad range of energies and dot-shapes. In
particular, this is the case for the ``diagonal contributions'' to
the above double sum (with $\gamma'=\gamma$), which can be
analyzed using the sum rule (in the spirit of Eq.~(B6) of  Ref.~\cite{Bar91})
 \begin{eqnarray}
 \sum_{\gamma} A_{\gamma}^2 \left[ \cdots \right]_{\gamma}  \! &=& \! \!
 \int_{-\frac{\pi}{2}}^{\frac{\pi}{2}} {\rm d} \theta_{0}
\int_{-\frac{\pi}{2}}^{\frac{\pi}{2}} {\rm d} \theta\,
  p_{\rm F} \cos \theta_0 \ 
\nonumber\\
& & \qquad \times \tilde{P} ({\bf Y} ,{\bf Y}_0; t)
  \left[ \cdots \right]_{{\bf Y}_0}.  \ \ 
  \label{Eq:sum-rule}
 \end{eqnarray}
Here we define ${\tilde P}({\bf Y},{\bf Y}_0;t)\de y\de \theta \de
t$ as the classical probability for a particle to go from an
initial position and momentum angle of ${\bf
Y}_0\equiv(y_0,\theta_0)$ on lead L to within $(\de y,\de \theta)$
of ${\bf Y}=(y,\theta)$ on lead R in a time within $\de t$ of $t$.
The average of ${\tilde P}$ over an ensemble of dots or over
energy results in a smooth function. If the dynamics are mixing on
a timescale $\ll \tau_{\rm D}$, one can approximate $ \left
\langle \tilde{P} ({\bf Y};{\bf Y}_0; t) \right\rangle =
e^{-t/\tau_{\rm D} } \cos \theta /[2 \left( W_{\rm L} + W_{\rm R}
\right) \tau_{\rm D}]$, which results in the classical Drude
conductance
\begin{eqnarray}
\langle g \rangle_{\rm D} = {N_{\rm L}N_{\rm R} \over (N_{\rm L} +
N_{\rm R})}. \label{eq:drude}
\end{eqnarray}
Quantum corrections to this result originate from correlations of
paths $\gamma$ and $\gamma'$ which are not identical, but closely
related by additional discrete symmetries in the system. For
asymmetric quantum dots the only possible additional symmetry is
time-reversal symmetry, which results in the ordinary
weak localization correction \cite{Aleiner-Larkin,Richter-Sieber,haake-conductance}
and associated coherent-backscattering peak 
\cite{Richter-Sieber,Jac06,Rahav-Brouwer-backscatter} 
for systems whose classical dynamics exhibit hyperbolic chaos.
The identification of possible pairings is also at the heart of
the calculation of the magnitude ${\rm var}(g)$ of universal
conductance fluctuations, which in the semiclassical theory
naturally takes the form of a quadruple sum over classical paths
\cite{Brouwer-Rahav-ucfs,Brouwer-Rahav-ucfs2}.

Spatial symmetries in such chaotic systems 
induce further possible pairings both for the
average conductance as well as for its variance, which are
discussed in detail in the first article in this series
\cite{WSK1}. In the following sections we revisit  these results
and extend them to the case of displaced leads, which is far
richer than the case of symmetry-respecting leads.

%=========================
\section{Left-right symmetric quantum dot with
displaced leads} \label{sect:LR}

%%%%%%%%%%%%%%%%%%%%%%%%%%%%%%%%%%%%%%%%%%%%%%%%%%%%%%%%%%%%%%%%%%%%%%%%%%%%%
\begin{figure*}
\centerline{
\hbox{
\includegraphics[width=0.7\textwidth]{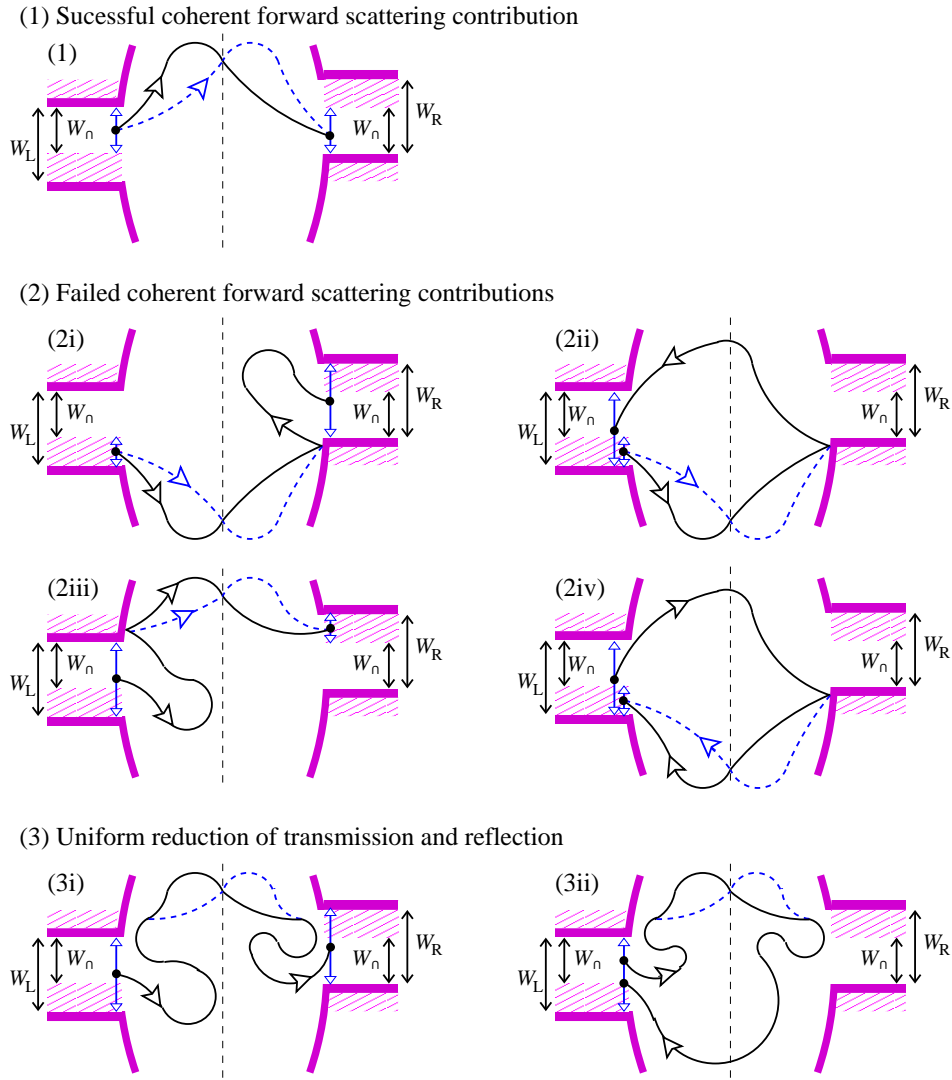}}}
\caption[]{\label{Fig:asymmleads-contributions} 
(colour online). List of
interference contributions to the conductance for a dot with
left-right mirror-symmetry when the leads are {\it asymmetric}.
Here the leads have widths $W_{\rm L}$ and $W_{\rm R}$ and are
centred at different places. The intersection of the L lead and
the R lead's mirror image has a width $W_\cap$ and is indicated by
the unshaded part of the L lead. The sketches on the left are all
contributions to transmission from the L lead to the R lead (hence
the contributions to conductance). The sketches on the right are
all contributions to reflection from the L lead back to the L
lead. }
\end{figure*}
%%%%%%%%%%%%%%%%%%%%%%%%%%%%%%%%%%%%%%%%%%%%%%%%%%%%%%%%%%%%%%%%%%%%%%%%%%%%%

\begin{figure}
\centerline{\hbox{
\includegraphics[width=1.0\columnwidth]{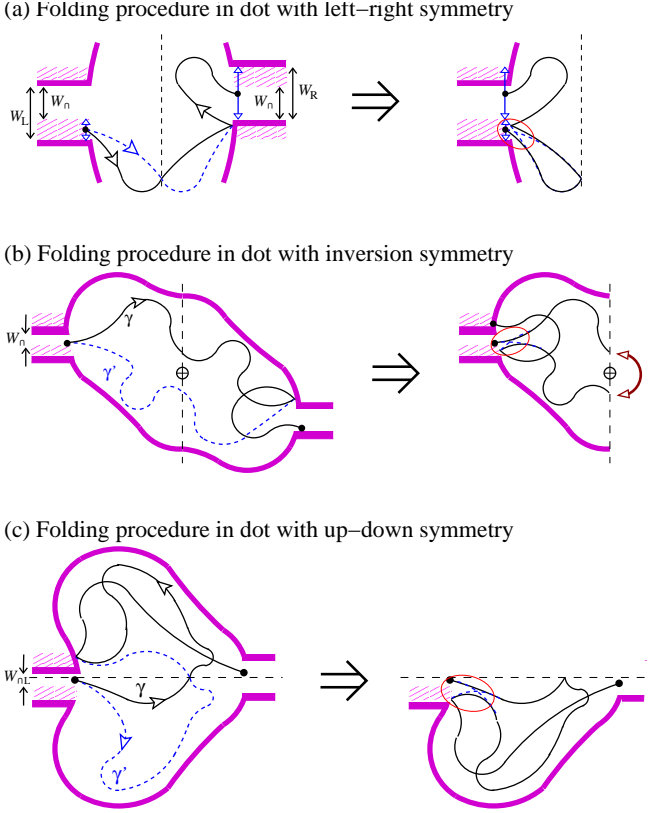}}}
\caption[]{\label{Fig:asymmleads-folds} 
(colour online). To find the non-trivial
path pairings, and to evaluate the phase difference between the
paths, we use the folding procedure introduced in Ref.~\cite{WSK1}
(for symmetric leads). Here we consider the extra contributions
generated by the fact the leads are asymmetric (i.e.,
contributions 2i-iv in Fig.~\ref{Fig:asymmleads-contributions}).
For each spatial symmetry, we give one example of the folding
procedure for an unsuccessful coherent forward-scattering (or
backscattering). The ellipses mark the effective encounters, where
paths interchange their pairing. The other contributions are
easily analyzed in the same way. }
\end{figure}

We first consider a left-right mirror-symmetric system with leads
that are (partially or fully) displaced from the
symmetry-respecting configuration. As shown in
Fig.~\ref{Fig:asymmleads}(a), the leads are of different widths
and centred at different places. The amount of symmetry-breaking
is characterized by the (possibly vanishing)  width $W_\cap$ of
intersection between lead L and the mirror image of lead R. In
Fig.~\ref{Fig:asymmleads-contributions} we show the path-pairings
for all symmetry-induced interference corrections to the average
conductance. (There is a strong resemblance between these
contributions and the weak localization correction for systems
with leads that contain tunnel barriers; in particular compare the
{\it failed coherent forward scattering} contributions in
Fig.~\ref{Fig:asymmleads-contributions} of this article with the
{\it failed coherent backscattering} contributions in Fig.~4 of
Ref.~\cite{Whi07}.) None of the contributions listed in
Fig.~\ref{Fig:asymmleads-contributions} are particularly difficult
to calculate using the method presented in the first of this pair
of article \cite{WSK1}. This method involves folding paths under
the spatial symmetry to find ways in which one can construct
pairings between paths or their images, with pairings switching at
``effective'' encounters; see Fig.~\ref{Fig:asymmleads-folds}. The
difficulty is to find all contributions. One crucial check is to
verify that the sum of all interference contributions to
transmission and reflection gives zero, thereby ensuring particle
conservation.

%%%%%%%%%%%%%%%%%%%%%%%%%%%%%%%%%%%%%%%%%%%%%%%%%%%%%%%%%%%%%%%%%%%%%%%%%%%%%
%%%%%%%%%%%%%%%%%%%%%%%%%%%%%%%%%%%%%%%%%%%%%%%%%%%%%%%%%%%%%%%%%%%%%%%%%%%%%

The main difference from the equivalent calculation for a system
with symmetric leads (cf. Ref.~\cite{WSK1}) is that here a pair of
symmetry-related paths has a shorter \emph{joint} survival time
than the pairs of identical paths in the diagonal contribution.
When the leads are symmetrically placed, the probability of a path
staying in the dot (not hitting a lead) is strictly identical to
the probability of its mirror image staying in the dot. 
This ceases to be the case
when the leads are not symmetric. We deal with this by explicitly
considering all situations where a path hits a lead (in which case
it escapes from the system) or the mirror image of a lead (in
which its mirror image will  escapes from the system). The
probability that either of the processes  occurs is $(W_{\rm
L}+W_{\rm R}-W_\cap)/C$ per bounce at the boundary of the dot,
where $C$ is the circumference of the dot. We therefore define a
modified dwell time
\begin{eqnarray}
\tD'
&=& \tD \times{W_{\rm L}+W_{\rm R} \over 2(W_{\rm L}+W_{\rm R}-W_\cap)}
\nonumber \\
&=& \tD \times{N_{\rm L}+N_{\rm R} \over 2(N_{\rm L}+N_{\rm R}-N_\cap)}
\end{eqnarray}
which characterizes the probability $\exp[-t/\tD']$ that a path
\emph{and} its mirror image are \emph{both} still in the dot at
time $t$. We use this probability in place of $\exp[-t/\tD]$ in
evaluating all parts of contributions 1 and 2 in
Fig.~\ref{Fig:asymmleads-contributions} where the paths are the
mirror image of each other.

\subsection{Successful and failed forward-scattering contributions}
\label{Sect:suc-and failed-forward}

The contribution of paths of the type labelled 1 and 2i-2iv in
Fig.~\ref{Fig:asymmleads-contributions} have an effective
encounter close to a lead.
 These contributions are
similar to certain contributions in an asymmetric system with
tunnel barriers \cite{Whi07}, and hence we use a similar method 
to analyze them here. The behavior of path $\gamma'$ is
completely determined by that of path $\gamma$, so the two paths
have the same amplitudes, $A_{\gamma'}=A_\gamma$. The action
difference between them is $(S_{\gamma}-S_{\gamma'}) =
(p_{0\perp}+m\Lambda r_{0\perp})r_{0\perp}$, where
$(r_{0\perp},p_{0\perp})$ is the component of $({\bf Y}-{\bf
Y}_0)$ which is perpendicular to the direction of path $\gamma$ at
${\bf Y}$ \cite{footnote:failed-cbs}. Using the sum rule in
Eq.~(\ref{Eq:sum-rule}), we see that the contribution 1 in
Fig.~\ref{Fig:asymmleads-contributions} is given by (cf.\
contribution LR:a in Ref.~\cite{WSK1})
\begin{eqnarray}
\label{eq:g_LR1}
\langle \de g \rangle_{\rm LR:1}
&=& (2\pi \hbar)^{-2}
\!\int_{\rm \cap} \!\! \rmd {\bf Y}_0 \!\int_{\rm \cap} \!\!  \rmd {\bf Y}
\int_0^\infty \!\! \rmd t
\\
& &\times p_{\rm F} \cos \theta_0 \,\langle P'({\bf
Y},{\bf Y}_0;t)\rangle \; {\rm Re}\big[\e^{\rmi
(S_{\gamma}-S_{\gamma'}) /\hbar}\big] .
\nonumber 
\end{eqnarray}
The limits on the integral indicate that we only integrate over
the region of the leads which have an overlap with each other
under the left-right mirror symmetry (the regions of width
$W_\cap$ marked in Fig.~\ref{Fig:asymmleads-contributions}).

The survival probability $\langle P'({\bf Y},{\bf Y}_0;t)\rangle =
\exp[-t/\tD'] \de r \de \theta /[\pi (W_{\rm L}+W_{\rm
R}-W_\cap)\tD']$ is that of a path and its mirror image. The
probability per unit time for path $\gamma$ to hit within $(\de
r,\de \theta)$ of a given point in the region of phase space
defined by the union of leads and their mirror images is $\langle
P'({\bf Y},{\bf Y}_0;t)\rangle \de {\bf Y}$ where $\de {\bf
Y}\equiv  \de r \de \theta$. Note that it is $\tD'$ rather than
$\tD$ which gives the decay rate of $\langle P'({\bf Y},{\bf
Y}_0;t)\rangle$. We express the ${\bf Y}_0$ integral in terms of
the relative coordinates $(r_{0\perp}, p_{0\perp})$ and define
$T'_{W}(r_{0\perp}, p_{0\perp})$ and $T'_{L}(r_{0\perp},
p_{0\perp})$ as the time between touching the lead and the
perpendicular distance between $\gamma$ and $\gamma'$ becoming of
order $W$ and $L$, respectively. For times less than
$T'_{W}(r_{0\perp}, p_{0\perp})$, the path segments are almost
mirror images of each other, and their joint survival probability
is the survival probability of a path and its mirror image. For
times longer than this the path-pairs escape independently, but
since the pairs are made of a path and its mirror image, the
escape rate is $\tD'$ not $\tD$. The $t$-integral in
Eq.~(\ref{eq:g_LR1}) must have a lower cut-off at
$2T'_L(r_{0\perp}, p_{0\perp})$, because that is the minimum time
for  reconvergence. (For shorter times there is no contribution, because path $\gamma$ and $\gamma'$ must 
separate to a distance of order the dot size, if they are going to reconverge
at the other lead). Thus we have
\begin{eqnarray}
& & \hskip -10mm
\int_{\cap} \! \rmd {\bf Y} \int_0^\infty \rmd t \langle
P'({\bf Y},{\bf Y}_0;t)\rangle 
\nonumber \\
&=& {N_\cap \exp[-T'_W/\tD'
-2(T'_L-T'_W)/\tD'] \over  N_{\rm L}+N_{\rm R}-N_\cap},
\label{Eq:int-of-P-asymleads}
\end{eqnarray}
where $T'_{L,W}$ are shorthand for $T'_{L,W}(r_{0\perp},
p_{0\perp})$.  Note that the $\tD'$ in the denominator of $\langle
P'({\bf Y},{\bf Y}_0;t)\rangle$ was cancelled when we integrated
over all times longer than $2T'_L$. For small $(p_{0\perp}+
m\Lambda r_{0\perp})$ we find
\begin{eqnarray}
T'_L(r_{0\perp}, p_{0\perp}) &\simeq& \Lambda^{-1} \ln \left[ {
m\Lambda L \over |p_{0\perp}+ m\Lambda r_{0\perp}| }\right] ,
\qquad \label{eq:T'_L}
\end{eqnarray}
and $T'_W(r_{0\perp}, p_{0\perp})$ is given by the same formula with
$L$ replaced by $W$. Evaluating the integrals over the relative
coordinates $(r_{0\perp}, p_{0\perp})$ as in Ref.~\cite{WSK1}, we
finally obtain
\begin{eqnarray}
\langle \de g\rangle_{\rm LR:1}
= N_\cap [2(N_{\rm L}+N_{\rm R}-N_\cap)]^{-1} \exp[-\tEc/\tD'].
\end{eqnarray}

The failed  coherent forward-scattering contributions labelled  2i
and 2ii in Fig.~\ref{Fig:asymmleads-contributions} come from the
window of width $W_{\rm L}-W_\cap$ in the L lead. This causes an
enhanced probability of hitting the mirror image of that part of
lead L. However, the lead R is not there, so this constructive
interference peak gets reflected back into the dot, and has a
probability of $N_{\rm R}/(N_{\rm L}+N_{\rm R})$ of going to lead
R and a probability of $N_{\rm L}/(N_{\rm L}+N_{\rm R})$ of going
back to lead L \cite{footnote:failed-cbs}. The former is a
contribution to transmission (and hence to the conductance) while
the latter is a contribution to reflection. Thus we have
\begin{eqnarray}
\langle \de g\rangle_{\rm LR:2i}
&=& {(N_{\rm L}-N_\cap) \,N_{\rm R} \, \exp[-\tEc/\tD'] 
\over 2(N_{\rm L}+N_{\rm R}-N_\cap)(N_{\rm L}+N_{\rm R})}  ,
\label{eq:g_LR:2i}
\\
\langle \de R\rangle_{\rm LR:2ii}
&=& 
{(N_{\rm L}-N_\cap) \,N_{\rm L} \, \exp[-\tEc/\tD'] 
\over 2(N_{\rm L}+N_{\rm R}-N_\cap)(N_{\rm L}+N_{\rm R})}.
\end{eqnarray}
By inspection of Fig.~\ref{Fig:asymmleads-contributions} it
follows that 
$\langle \de g\rangle_{\rm LR:2iii}$ 
is given by Eq.~(\ref{eq:g_LR:2i}) with $N_{\rm R}$ and $N_{\rm L}$
interchanged,
while 
$\langle \de R\rangle_{\rm LR:2iv}=\langle \de R\rangle_{\rm LR:2ii}$.

\subsection{Uniform contributions to  transmission and reflection}
\label{Sect:uniform-contrib}

To evaluate the uniform contributions to transmission and
reflection, labelled 3i and 3ii in
Fig.~\ref{Fig:asymmleads-contributions}, we divide  the pairs of
paths in this contribution  into three regions. The first part is
when $\gamma$ and $\gamma'$ are the same, and are far from the
encounter (a time $T_W/2$ or more away from the encounter). Here
the probability for the paths to escape is $1/\tD$ per unit time.
The second region is where $\gamma'$ and $\gamma$ are the mirror
image of each other and far from the encounter (a time $T_W/2$ or
more from the encounter). Here the probability of one or both
paths to escape is $1/\tD'$ per unit time. Finally, the third
region is close to the encounter (less than a time $T_W/2$ away
from the encounter). Here the probability for the paths to escape
the first time they pass through this region surrounding the
encounter is $\exp[-T_W/\tD]$. However, the conditional
probability to escape the second time the paths pass through this
region
 (given that they both survived the first time) is
$\exp\big[ -T_W \big( 1/\tD'-1/\tD \big) \big]$. It follows that
the contribution 3i is given by $ \langle \de g \rangle_{\rm
LR:3i} = (\pi \hbar)^{-1} \int_{\rm L} \! {\rm d} {\bf Y}_{0}\int
{\rm d} \epsilon \,
 {\rm Re}\big[e^{i(S_{\gamma}-S_{\gamma'}) /\hbar }\big]
 \big\langle  F ( {\bf Y}_{0}, \epsilon)  \big\rangle$,
where the action difference $(S_\gamma -S_{\gamma'})$ is the same as 
for weak localization in Refs.~\cite{Sie01,Richter-Sieber} 
and
 \begin{eqnarray}
 F( {\bf Y}_{0}, \epsilon)  &=&
 2v_{\rm F}^2  \sin \epsilon
  \int_{T_L+T_W} ^{\infty} \!  {\rm d } t
 \int_{T_ L+\frac{T_W}{2}}^{t-\frac{T_W }{2} } \!   {\rm d} t_2
 \int_{\frac{T_W}{2}}^{t_2-T_L}  \!   {\rm d} t_1
\nonumber\\
&\times&
 p_{\rm F}\cos \theta_0 \int_{\rm R} \!  {\rm d} {\bf Y}
 \int_{ C} \! {\rm d} {\bf R}_1
 \tilde{P}({\bf Y} ,{\bf R}_2 ;t-t_2)
 \nonumber \\
&\times&
\tilde{P}'({\bf R}_2,{\bf R}_1;t_2-t_1)
 {\tilde P}({\bf R}_1,{\bf Y}_0;t_1).
\label{Eq:F}
 \end{eqnarray}
Since the paths are paired with their mirror image between time
$t_1$ and time $t_2$, the survival rate is $\tD'$ during this
time, but it is $\tD$ at all other times. Evaluating this integral
with these survival times gives
\begin{eqnarray}
\left\langle  F( {\bf Y}_{0}, \epsilon)  \right\rangle
&=&  {2v_{\rm F}^2\tD \tD' \over 2\pi A}
{N_{\rm R} \over  N_{\rm L}+N_{\rm R}}
p_{\rm F} \cos\theta_0
\nonumber \\
& & \times \sin \eps \,\exp \big[-T_L(\eps)/\tD'\big].
\label{Eq:F-result-asym}
\end{eqnarray}
This has two differences from the result for symmetric leads in
Ref.~\cite{WSK1}.  The exponent contains $\tD'$ not $\tD$, and the
prefactor contains $\tD\tD'$ not $\tD^2$. When integrating over
$\eps$, we obtain a factor of $[\Lambda
\tD']^{-1}\exp[-\tEc/\tD']$ in place of $[\Lambda \tD]^{-1}
\exp[-\tEc/\tD]$. Thus the $\tD'$ in the prefactor is cancelled
\cite{footnote:tunnel-wl}. Evaluating the integrals, we get
\begin{eqnarray}
\langle \de g\rangle_{\rm LR:3i}
&=& - N_{\rm L}N_{\rm R} [N_{\rm L}+N_{\rm R}]^{-2} \exp[-\tEc/\tD'], \quad \ 
\\
\langle \de R\rangle_{\rm LR:3ii} &=& - N_{\rm L}^2 [N_{\rm
L}+N_{\rm R}]^{-2} \exp[-\tEc/\tD'].
\end{eqnarray}
%since the formula is found by
% $\langle \de R\rangle_{\rm LR:3ii}$ is the same
%as $\langle \de g\rangle_{\rm LR:3i}$
%with $N_{\rm L}$ in place of $N_{\rm R}$ in the numerator.
These results are of the same form as the weak localization
correction except that the exponent contains $\tD'$ in place of
$\tD$. In particular, we recover the familiar factor of $-N_{\rm
L}N_{\rm R}/(N_{\rm L}+N_{\rm R})^2$ even though the joint survival
time is reduced when the paths are mirror images of each other.

One can next include other suppression effects such as asymmetry
in the dot and dephasing, which we discussed for dots with
symmetric leads in Ref.~\cite{WSK1}. The only difference caused by
asymmetric leads is that now the parts of contributions affected
by asymmetries and dephasing (parts where paths are paired with
their mirror image) decay with a rate $\tD'$ instead of $\tD$.
Thus we find that all the contributions listed in
Fig.~\ref{Fig:asymmleads-contributions} are then multiplied by a
factor
\begin{eqnarray}
Z'_{\rm LR}(\gamma_{\rm asym},\gamma_\phi) = {\exp[
-\gamma_\phi\tilde{\tau}-\gamma_{\rm asym}\tilde{\tau}_{\rm asym}]
\over 1+(\gamma_{\rm asym}+\gamma_\phi)\tD'}, \label{Eq:Z'_LR}
\end{eqnarray}
where the expression for the decay rates $\gamma_{\rm asym}$,
$\gamma_\phi$ and timescales $\tilde{\tau}$, $\tilde{\tau}_{\rm
asym}$ are the same as for a dot with symmetric leads \cite{WSK1}.

\subsection{Conductance of a left-right symmetric quantum dot with asymmetric leads}

As required by particle number conservation, the seven
contributions in Fig.~\ref{Fig:asymmleads-contributions} sum to
zero. In order to obtain the conductance, we sum the four
contributions to transmission from the left lead to the right lead
(contributions 1, 2i, 2iii and 3i), and add them to the Drude
conductance and the weak localization correction. This gives the
conductance of a chaotic left-right symmetric  quantum dot with
many-modes on each lead ($N_{\rm L},N_{\rm R},N_\cap \gg 1$),
\begin{eqnarray}
\label{Eq:g-LR-final-result}
\langle g\rangle_{\rm LR}
&=& {N_{\rm L} N_{\rm R}\over N_{\rm L} +N_{\rm R}}
 \\
& & \hskip -3mm + {N_{\rm L}N_{\rm R}
\over (N_{\rm L}+N_{\rm R})^2}
\Bigg[ {N_\cap \e^{-\tEc/\tD'} \over N_{\rm L}+N_{\rm R}-N_\cap}
Z'_{\rm LR}(\gamma_{\rm asym},\gamma_\phi)
\nonumber \\
& &\qquad - \e^{-\tEc/\tD} Z(B,\gamma_\phi) \Bigg]
 +{\cal O}[N_{\rm L,R}^{-1}],
\nonumber
\end{eqnarray}
where $Z'_{\rm LR} (\gamma_{\rm asym},\gamma_\phi)$ is given by
Eq.~(\ref{Eq:Z'_LR}).  The second term in the square brackets is
the usual weak localization correction, which is suppressed by
magnetic fields and dephasing according to the function
$Z(B,\gamma_\phi) = \exp[-\gamma_\phi\tilde{\tau}]
\big[1+ (B/B_c)^2+\gamma_\phi\tau_{\rm
D}\big]^{-1}$.
For symmetric leads we have $N_\cap=N_{\rm L}=N_{\rm R}$ (and
hence $\tD'=\tD$), and this result immediately reduces to the one
in Ref.~\cite{WSK1}.

It is worth considering two special cases. The first case is when
the leads are of equal width but not centred at the mirror image
of each other, such that $N_\cap < N_{\rm L}=N_{\rm R}\equiv N$.
Taking $\De w=1-w_\cap/W= 1 -N_\cap/N$ as the relative distance
(in units of the lead width $W=W_{\rm L}=W_{\rm R}$) by which lead
L is displaced with respect to the mirror image of lead R, and
assuming there is no dephasing, magnetic field, or internal
asymmetry, we find
\begin{eqnarray}
\langle g\rangle_{\rm LR} &=& {N\over 2} + {1 \over 4} \left[
{1-\De w \over 1+\De w} \e^{-\tEc/\tD'} - \e^{-\tEc/\tD} \right] 
\nonumber \\
& & +
{\cal O}[N^{-1}]. \label{Eq:move-lead}
\end{eqnarray}

The second special case is when the lead R is narrower but
situated entirely within the mirror image of lead L;  we then have
$N_\cap=N_{\rm R} < N_{\rm L}$. Assuming again that there is no
dephasing, magnetic field, or internal asymmetry,
\begin{eqnarray}
\langle g\rangle_{\rm LR}
&=& {N_{\rm L} N_{\rm R}\over N_{\rm L} +N_{\rm R}}
\nonumber \\
& &
+ {N_{\rm L}N_{\rm R}
\over (N_{\rm L}+N_{\rm R})^2}
\left[ {N_{\rm R} \over N_{\rm L}} \e^{-\tEc/\tD'}
 - \e^{-\tEc/\tD} \right]
\nonumber \\
& & +\ {\cal O}[N_{\rm L,R}^{-1}]. \label{Eq:narrow-Rlead}
\end{eqnarray}
As one could scan the narrow lead R across the mirror image of the
wide lead L, this scenario can be thought of as a probe of the
shape of the coherent forward-scattering peak.  The fact that our
result Eq.~(\ref{Eq:narrow-Rlead}) is independent of the position
of lead R tells us that the forward-scattering peak is uniformly
distributed over the region defined by the mirror image of lead L.

%=================================
\section{Inversion-symmetric quantum dot with asymmetric leads}
\label{sect:inv}

For systems with inversion symmetry the calculation follows much
as for a left-right symmetry. The one significant difference is
the magnetic-field dependence of the contributions, which was
treated in Ref.~\cite{WSK1}. The displacement of the leads simply
requires us to replace $\tD$ with $\tD'$ in the suppression of
contributions by magnetic fields, asymmetries in the dot, and
dephasing. The suppression factor therefore takes the form
\begin{eqnarray}
Z'_{\rm inv}(B,\gamma_{\rm asym},\gamma_\phi) = {\exp[-\gamma_{\rm
asym}\tilde\tau_{\rm asym} -\gamma_\phi\tilde{\tau}] \over 1+
(B/B'_c)^2 +(\gamma_{\rm
asym}+\gamma_\phi)\tD'}. 
\nonumber \\
\label{Eq:Z'_inv}
\end{eqnarray}
where $B'_c= aB_0\sqrt{\tau_0/2\tD'}$ is given by Eq.~(\ref{eq:bc}) 
with $\tD$ replaced by $\tD'$. 
As a result, an inversion-symmetric quantum dot with many modes on
each lead ($N_{\rm L},N_{\rm R},N_\cap \gg 1$) has a total average
conductance of
\begin{eqnarray}
\langle g\rangle_{\rm inv}
&=& {N_{\rm L} N_{\rm R}\over N_{\rm L} +N_{\rm R}}
\nonumber \\
&+& \! \! {N_{\rm L}N_{\rm R}
\over (N_{\rm L}+N_{\rm R})^2}
\Bigg[ {N_\cap \e^{-\tEc/\tD'} \over N_{\rm L}+N_{\rm R}-N_\cap}
Z'_{\rm inv}(B,\gamma_{\rm asym},\gamma_\phi)
\nonumber \\
& & - \e^{-\tEc/\tD} Z_{\rm wl}(B,\gamma_\phi) \Bigg]
+ {\cal O}[N^{-1}_{\rm L,R}].
\end{eqnarray}
With the exception of the magnetic-field dependence of the second
term, this formula is the same as Eq.\
(\ref{Eq:g-LR-final-result}) for a left-right symmetric dot. Thus
the two special cases discussed below
Eq.~(\ref{Eq:g-LR-final-result}) are directly applicable here.

\section{Up-down symmetric quantum dot with asymmetric leads}
\label{sect:UD}

For up-down symmetric systems, there are a number of important
differences with the case of left-right symmetry discussed in
Section~\ref{sect:LR}. Firstly, a pair of paths related by the
mirror symmetry decays jointly at a rate
\begin{eqnarray}
\tD^{\rm (UD)} = \tD \times {N_{\rm L}+N_{\rm R} \over 2N_{\rm
L}+2N_{\rm R}-N_{\rm \cap L}-N_{\rm \cap R}},
\label{eq:tau-updown}
\end{eqnarray}
where $N_{\rm \cap L}$ is the number of modes in the intersection
of lead L with its own mirror image, and $N_{\rm \cap R}$ is the
number of modes in the intersection of lead R with its own mirror
image. Secondly, the successful and failed forward-scattering
contributions for left-right symmetry are converted into
successful and failed backscattering contributions for up-down
symmetry. In  particular, successful backscattering makes no
contribute to the conductance. The other contributions to
transmission are not very different from those for left-right
symmetry, except that one must distinguish $N_{\rm \cap L}$ from
$N_{\rm \cap R}$, and one must replace $\tD'$ by $\tD^{\rm (UD)}$.
Summing up the contributions to conductance induced by the spatial
symmetry, we find
\begin{eqnarray}
\langle \de g\rangle_{\rm UD} &=& - {(N_{\rm \cap L} N_{\rm R}^2
+ N_{\rm \cap R} N_{\rm L}^2) \exp [-\tEc/\tD^{\rm (UD)}] \over
(2N_{\rm L}+2N_{\rm R}-N_{\rm \cap L}-N_{\rm \cap R}) (N_{\rm
L}+N_{\rm R})^2} 
\nonumber \\
& &\times Z'_{\rm UD} (\gamma_{\rm
asym},\gamma_\phi), \label{Eq:de-g_UD}
\end{eqnarray}
where $Z'_{\rm UD}(\gamma_{\rm asym},\gamma_\phi)$ has the same
form as $Z'_{\rm LR}(\gamma_{\rm asym},\gamma_\phi)$ given in
Eq.~(\ref{Eq:Z'_LR}), but with $\tD'$ replaced by $\tD^{\rm
(UD)}$. Like for left-right mirror-symmetry (but unlike for 
inversion symmetry) this contribution is unaffected by a 
magnetic field.

\begin{widetext}
The average conductance of an up-down mirror-symmetric dot
with many modes on each lead is therefore
\begin{eqnarray}
\langle g\rangle_{\rm UD}
&=& {N_{\rm L}N_{\rm R} \over N_{\rm L}+N_{\rm R}}
% \nonumber \\ & & 
-{N_{\rm L}N_{\rm R} \over (N_{\rm L}+N_{\rm R})^2}
\Bigg[ \left( {N_{\rm \cap L}N_{\rm R}\over N_{\rm L}} +
{N_{\rm \cap R}N_{\rm L}\over N_{\rm R}}\right)
{\exp[{-\tEc/\tD^{\rm (UD)}}] \ Z'_{\rm UD} (\gamma_{\rm asym},\gamma_\phi)
\over 2N_{\rm L}+2N_{\rm R}-N_{\rm \cap L}-N_{\rm \cap R} }
\nonumber \\
& &  \qquad \qquad \qquad \qquad + \exp[{-\tEc/\tD}] \ Z_{\rm wl}
(B,\gamma_\phi) \Bigg] \ + \ {\cal O}[N_{\rm L,R}^{-1}]. \label{Eq:g_UD}
\end{eqnarray}
\end{widetext}

It is worth noting that the spatial symmetry induces a reduction
of conductance whenever one lead is close to symmetric, even if
the other lead is completely asymmetric (i.\,e. when $N_{\rm \cap
L}=0$ but $N_{\rm \cap R} \neq 0$, or vice versa). For example,
when both leads have the same width ($N_{\rm L} = N_{\rm R}=N$)
and the right lead is perfectly on the symmetry axis ($N_{\rm \cap
R} = N_{\rm R}$), but the left lead is a long way from the
symmetry axis ($N_{\rm \cap L} = 0$), Eq.~(\ref{Eq:g_UD}) reduces
to
\begin{eqnarray}
\langle g\rangle_{\rm UD} &=& {N\over 2} - {1 \over 4} \Bigg[
{1 \over 3} \exp[{-\tEc/\tD^{\rm (UD)}}] + \exp[{-\tEc/\tD}]
\Bigg] 
\nonumber \\
& & \qquad \qquad + \ {\cal O}[N^{-1}]
\end{eqnarray}
assuming no dephasing, magnetic field and no asymmetry in the dot. 
If the Ehrenfest time is much shorter than $\tD$ and $\tD^{\rm UD}$, the
average conductance of the system with one displaced lead is
therefore simply $\langle g\rangle_{\rm UD} = N/2-1/3$.

Remarkably, the conductance from the L lead to the R lead is
therefore affected by the symmetry of the dot even when the L lead
is completely asymmetric. this result is perhaps less
counterintuitive when one considers reflection (rather than
transmission). If one lead is on the symmetry axis, then
reflection back to that lead will be enhanced even if the other
lead is a long way from the symmetry axis. Since we have particle
conservation, there must be an associated reduction in
transmission from one lead to the other (compared to transmission
in a completely asymmetric situation).

%============================
\section{Four-fold symmetric   quantum dot with asymmetric leads}
\label{sect:4F}

A quantum dot with four-fold symmetry simultaneously possesses all
three of the spatial symmetries that we discuss in this article.
The interference corrections to the conductance of such a system
are simply the sum of the corrections due to each of these three
symmetries (i.e., the presence of the extra symmetries has no
effect on the contributions which do not respect those
symmetries),
\begin{eqnarray}
\label{Eq:g4F} 
\langle \de g\rangle_{\rm 4F} &=& \langle \de
g\rangle_{\rm LR} + \langle \de g\rangle_{\rm inv} +\langle \de
g\rangle_{\rm UD},
\end{eqnarray}
where $\langle\de g\rangle_\kappa$ is the contribution to the
average conductance induced by spatial symmetry
 $\kappa \in {\rm LR},{\rm inv},{\rm UD}$. The explicit form of this result is
easily extracted from the expressions in the previous sections.
Instead of writing it out in full, we consider the special
case where the two leads have the same width, $N_{\rm L} = N_{\rm
R}=N$, the Ehrenfest time is negligible and there is no dephasing,
magnetic field or asymmetry in the dot. The average conductance
then takes the form
\begin{eqnarray}
\langle g\rangle_{\rm 4F} &=& {N \over 2} + {1\over 4}\Big[
{N_{\rm \cap LR} \over 2N-N_{\rm \cap LR}} + {N_{\rm \cap inv}
\over 2N-N_{\rm \cap inv}} 
\nonumber \\
& & \qquad - {N_{\rm \cap UD:L} + N_{\rm \cap
UD:R} \over 4N-N_{\rm \cap UD:L}-N_{\rm \cap UD:R} } -1 \Big],
\label{Eq:delta-g-4fold}
\end{eqnarray}
where $N_{\rm \cap LR}$ is the intersection between leads  L and R
under the left-right symmetry, $N_{\rm \cap inv}$ is the
intersection between leads L and R under the inversion symmetry,
and $N_{\rm \cap UD:L}$ ($N_{\rm \cap UD:R}$) is the intersection
of lead L (R)  with itself under the up-down symmetry. The final
term in the square-bracket is the usual weak localization
contribution.

Since the presence of two of the above mentioned symmetries always
implies the presence of the third, it is not possible to move the
leads such that only one of the $N_\cap$ parameters changes.
Without affecting the integrity of the leads there are only two
possible modifications for which only two of the parameters
change; starting with perfectly symmetric leads one can (a) move
both leads upwards by the same amount so that $N_{\rm \cap LR}$ is
unchanged, or (b) move both leads by the same amount in opposite
directions (one up and one down) so that $N_{\rm \cap inv}$ is
unchanged. In principle, it is also possible to break up a single
lead (say L) in the middle and move the two parts into opposite
directions (both parts would still be contacted by the same source
or drain electrode); this preserves $N_{\rm \cap L}$ and $N_{\rm
\cap R}$ but affects the other parameters. However, the latter
deformation is difficult to realize in practice.

%============================
\section{Universal conductance fluctuations with displaced leads}
\label{sect:UCFs}

%%%%%%%%%%%%%%%%%%%%%%%%%%%%%%%%%%%%%%%%%%%%%%%%%%%%%%%%%%%%%%%%%%%%%%%%%%%%%
\begin{figure*}
\centerline{\hbox{
\includegraphics[width=0.7\textwidth]{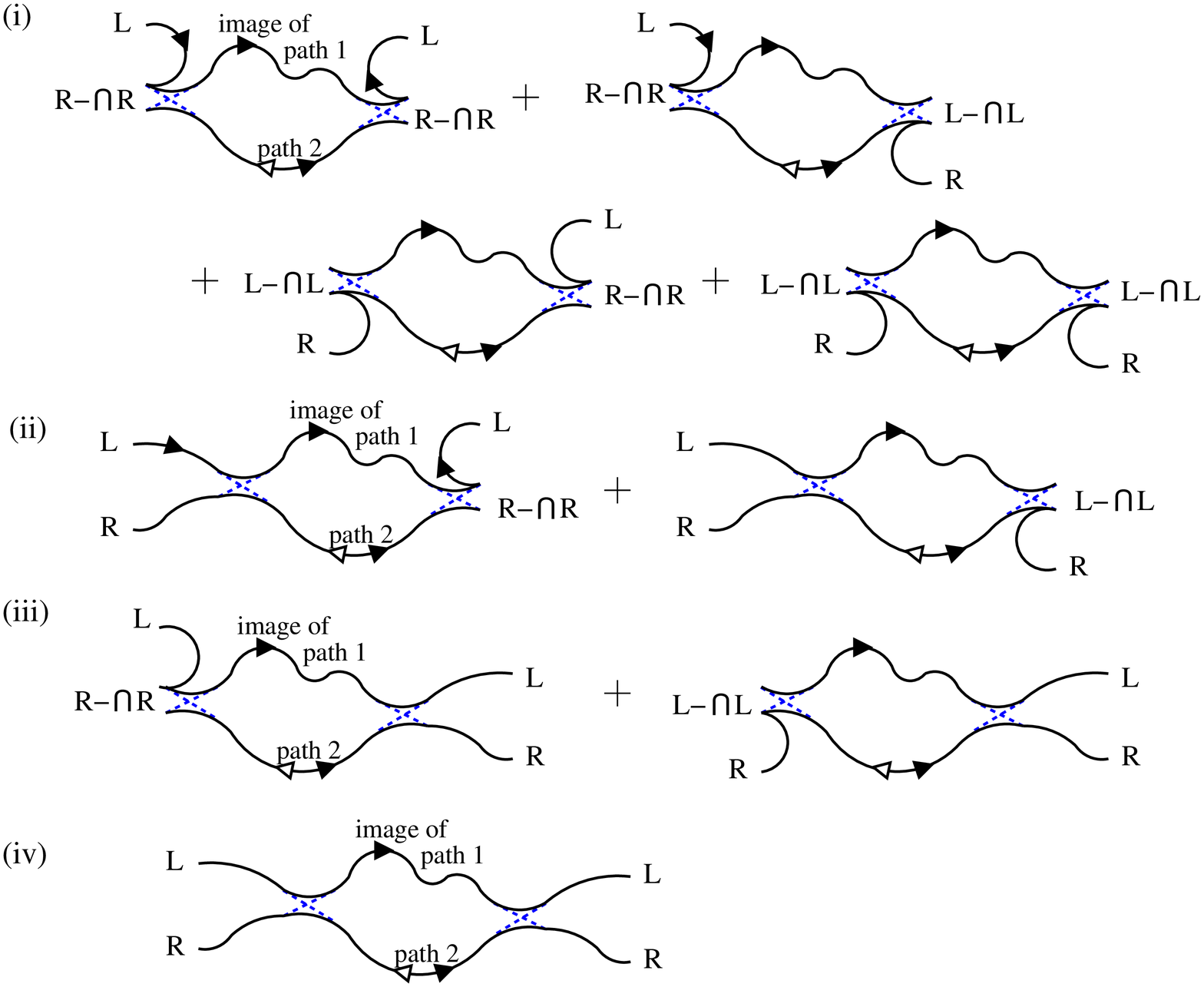}}}
\caption[]{\label{Fig:ucf-asymleads1} 
(colour online). A sketch of semiclassical
contributions to UCFs (more specifically, contributions to ${\rm
covar}[R,R']$) for an up-down symmetric dot with asymmetric leads.
There are analogous contributions to UCFs for left-right or
inversion-symmetric dots (see explanation in the text). In each
contribution, paths 1 and 1' go from L lead to L lead, while paths
2 and 2' go from R lead to R lead. In the sketches, solid lines
indicate paths 2 and the image (mirror image or image under the
inversion symmetry) of paths 1.  Path 2' and the image of path 1'
are indicated by the dashed lines (only shown at the encounters).
Thus when paths 2 and 2' are not paired with each other they are
paired with the image of 1' and 1 respectively (indicated by solid
arrowheads). If the system has a time-reversal symmetry then path
2 and 2' can also be paired with the time-reverses of the image of
1' and 1, respectively (indicated by the open arrowheads). }
\end{figure*}
%%%%%%%%%%%%%%%%%%%%%%%%%%%%%%%%%%%%%%%%%%%%%%%%%%%%%%%%%%%%%%%%%%%%%%%%%%%%%

%%%%%%%%%%%%%%%%%%%%%%%%%%%%%%%%%%%%%%%%%%%%%%%%%%%%%%%%%%%%%%%%%%%%%%%%%%%%%
\begin{figure*}
\centerline{\hbox{
\includegraphics[width=0.7\textwidth]{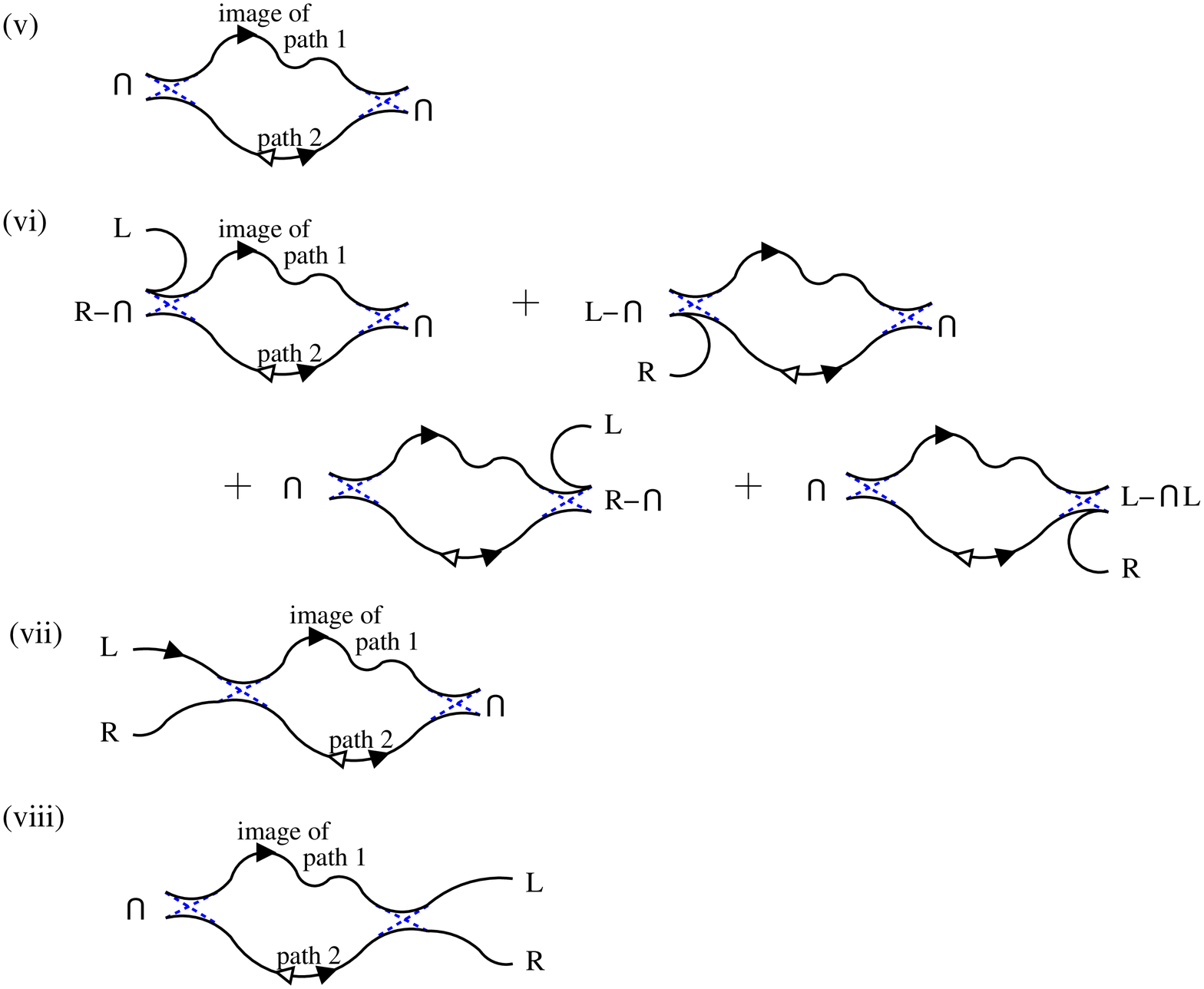}}}
\caption[]{\label{Fig:ucf-asymleads2} 
(colour online). A sketch of additional
semiclassical contributions to UCFs (more specifically,
contributions to ${\rm covar}[R,R']$) for left-right or
inversion-symmetric dots with asymmetric leads. The contributions
listed here must be {\it added} to those listed in
Fig.~\ref{Fig:ucf-asymleads1} (once one sets $\cap L=\cap R =
\cap$) to get the full set of contributions for left-right or
inversion-symmetric dots. The manner in which the contributions
are sketched is explained in the caption of
Fig.~\ref{Fig:ucf-asymleads1}. }
\end{figure*}
%%%%%%%%%%%%%%%%%%%%%%%%%%%%%%%%%%%%%%%%%%%%%%%%%%%%%%%%%%%%%%%%%%%%%%%%%%%%%

Now we turn to the magnitude of universal conductance fluctuations
(UCFs) in symmetric dots with asymmetric leads. Their calculation
is generally far more complicated than the calculation of the
average conductance. This is illustrated by the fact that there is
as yet no semiclassical theory of UCFs for leads with tunnel
barriers, a problem which has many similarities to the problem we
need to solve here. Thus we restrict ourselves to the simplest
case of quantum dots with negligible Ehrenfest time and negligible
dephasing, and only consider magnetic fields which are either
negligibly small ($B\ll B_c$), or sufficiently strong to break
time-reversal symmetry in the asymmetric system ($B\gg B_c$).

The magnitude of the UCFs (with conductances measured in units of $e^2/h$)
is given by ${\rm
var}[g]= {\rm var}[T]$, where $T=\tr[ t^\dagger t]$ and $t$ is the
block of the scattering matrix
$S=\left(%
\begin{array}{cc}
  r & t' \\
  t & r' \\
\end{array}%
\right)$ associated with transmission from lead L to lead R. For
practical calculations it is beneficial  to exploit the unitarity
of the scattering matrix (i.e., current conservation), which
results in the relations $T = N_{\rm L}- R= N_{\rm R}-R'$ with
$R=\tr[r^\dagger r]$ and $R'=\tr[r^{\prime\dagger} r^\prime]$,
where $r$ is the block of the scattering matrix associated with
reflection back to lead L, and $r^\prime$ describes reflection
back to lead R. As a result we can write the magnitude of the
UCFs in any of the following ways,
\begin{eqnarray}
{\rm var}[g] = {\rm var}[R] =  {\rm var}[R'] = {\rm covar}[R,R'].
\end{eqnarray}
As for conventional UCFs without spatial symmetries
\cite{Brouwer-Rahav-ucfs,Brouwer-Rahav-ucfs2}, the semiclassical
calculation of ${\rm covar}[R,R']$ is most straight-forward, thus
we base our calculations on this quantity. For the expert reader,
Appendix \ref{appendix} contains an outline of the calculation of ${\rm
var}[R]$ and ${\rm var}[R']$, showing that they equal ${\rm
covar}[R,R']$.

All symmetry-induced contributions to ${\rm covar}[R,R']$ for an
up-down symmetric dot are listed in Fig.~\ref{Fig:ucf-asymleads1}.
For a left-right or inversion-symmetric dot there are
\emph{additional} contributions, which are listed in
Fig.~\ref{Fig:ucf-asymleads2}. In all cases, when paths 2 and 2'
are not paired with each other, they are paired with the
\emph{images} of paths 1' and 1 under the appropriate symmetry
operation. To keep the sketches in Figs.~\ref{Fig:ucf-asymleads1}
and \ref{Fig:ucf-asymleads2} as clear as possible,  we  only show
these images of paths 1 and 1' (rather than paths 1 and 1'
themselves). Then the resulting contributions look very much like
the usual contributions to UCFs in a system without a spatial
symmetry \cite{Brouwer-Rahav-ucfs,Brouwer-Rahav-ucfs2}.

In analogy to the situation in asymmetric systems,
 one would also expect
contributions in which paths wind around periodic orbits (see
Figs.~1b,c in Ref.~\cite{Brouwer-Rahav-ucfs2}). For example, a
symmetric quantum dot will have contributions in which path 1' is
the same as path 1 except that it winds around a periodic orbit
$p$ when path 1 does not (thus path 1 must come very close to the
periodic orbit in phase space), while path 2 is the same as path
2' except that it winds around the {\it image}  of the orbit $p$.
These contributions are proportional to those analyzed for UCFs in
asymmetric dots, the only modification being that the joint
survival probability of a periodic orbit and its image is again
changed to $\exp[-t/\tau'_{\rm D}]$. Drawing on the results of
Refs.~\cite{Brouwer-Rahav-ucfs,Brouwer-Rahav-ucfs2}, it follows
that the contributions involving windings around periodic orbits
will be negligibly small  when the Ehrenfest time is small. (This
observation makes the calculation of the UCFs in the present
problem significantly simpler than for the case with tunnel
barriers, where one cannot rule out contributions from periodic
orbits which touch the barriers on the leads.)

\subsection{Effect of time-reversal symmetry}

%% The following means that we do not study the cross-over
Inspecting the sketches in Figs.\ \ref{Fig:ucf-asymleads1} and
\ref{Fig:ucf-asymleads2}  we see that all contributions are
doubled when the magnetic field is negligible, because path 2 can
either follow the image of path 1' or the time-reverse of path 1'.
Thus we can multiple all terms by $2/\beta$, where $\beta=1$ for a
system with negligible magnetic field, $B\ll B_c$ and $\beta=2$
for a system with a finite magnetic field, $B\gg B_{\rm c}$. In
the latter case the presence of mirror-reflection symmetries 
allows one to define a generalized time-reversal symmetry; however,
this is already accounted for in the construction of all diagrams
(see Appendix A of Ref.~\cite{WSK1}).

\subsection{UCFs in an up-down symmetric dot}

The general rules for constructing all contributions to the UCFs
are the following. Each segment where path 2 or 2' is paired with
the image of path 1' or 1 gives a factor of $(2N_{\rm L}+2N_{\rm
R}-N_{\rm \cap L}-N_{\rm \cap R})^{-1}$, which arises from the
survival time $\tD^{\rm UD}$ given in Eq.~(\ref{eq:tau-updown}).
Each segment where  paths 2 and 2' are paired (or paths 1 and 1'
are paired)
 gives a factor of $(N_{\rm L}+N_{\rm R})^{-1}$,
which comes from the conventional survival time $\tD$. Each
segment that touches a lead gives a factor equal to the number of
lead modes that the path could couple to; i.e., a lead labelled
``$R-\cap R$'' gives a factor of $(N_{\rm R}-N_{\rm \cap R})$,
while a lead labelled ``R'' simply gives a factor of $N_{\rm R}$.
An encounter which touches a lead gives the same factor as a
simple path-segment that touches a lead, so again if it is
labelled ``$R-\cap R$'' then it gives a factor of $(N_{\rm
R}-N_{\rm \cap R})$ (this rule is proven by applying the same
analysis as was used for the successful and failed
forward-scattering processes in Section~\ref{Sect:suc-and
failed-forward}.) Finally, encounters deep in the dot (i.e., those
which do not touch the leads) give a factor of $-(2N_{\rm
L}+2N_{\rm R}-N_{\rm \cap L}-N_{\rm \cap R})$ (this rule can be
proven by applying the same analysis as was used for the uniform
contributions to transmission in
Section~\ref{Sect:uniform-contrib}). With this set of rules we can
easily see that contribution (i) in Fig.\ \ref{Fig:ucf-asymleads1}
gives
\begin{widetext}
\begin{eqnarray}
C_{\rm i} &=& {2 \over \beta}\, {N_{\rm L}^2 (N_{\rm R}-N_{\rm
\cap R})^2 + 2 N_{\rm L}  (N_{\rm L}-N_{\rm \cap L})N_{\rm R}
(N_{\rm R}-N_{\rm \cap R}) + (N_{\rm L}-N_{\rm \cap L})^2N_{\rm
R}^2 \over (2N_{\rm L}+2N_{\rm R}-N_{\rm \cap L} -N_{\rm \cap
R})^2 (N_{\rm L}+N_{\rm R})^2}. \label{Eq:Ci}
\end{eqnarray}
Next we see that $C_{\rm iii}=C_{\rm ii}$, and that they are
negative because only one of the encounters is deep in the dot 
(the other is near a lead), resulting
in
\begin{eqnarray}
C_{\rm ii} + C_{\rm iii} &=& - 2\,{2 \over \beta}\, {N_{\rm L}^2
N_{\rm R} (N_{\rm R}-N_{\rm \cap R}) + N_{\rm L}  (N_{\rm
L}-N_{\rm \cap L})N_{\rm R}^2 \over (2N_{\rm L}+2N_{\rm R}-N_{\rm
\cap L} -N_{\rm \cap R})(N_{\rm L}+N_{\rm R})^3}. \label{Eq:Cii}
\end{eqnarray}
\end{widetext}
Finally $C_{\rm iv}$ gives a positive contribution because it has
two encounters deep in the dot, and is given by
\begin{eqnarray}
C_{\rm iv} &=&  {2 \over \beta} \, { N_{\rm L}^2N_{\rm R}^2  \over
(N_{\rm L}+N_{\rm R})^4}. \label{Eq:Civ}
\end{eqnarray}

The total magnitude of the UCFs is given by the UCFs of an
asymmetric dot, ${\rm var}[g]_{\rm asym}$, plus the sum of the
terms above, i.e., ${\rm var}[g] = {\rm var}[g]_{\rm asym}+C_{\rm
i}+C_{\rm ii}+C_{\rm iii}+C_{\rm iv}$. In the limit of perfectly
symmetric leads ($N_{\rm \cap L}=N_{\rm L}$ and $N_{\rm \cap
R}=N_{\rm R}$), only $C_{\rm iv}$ survives and the UCFs have
double the magnitude as those for an asymmetric dot. In the limit
of completely asymmetric leads ($N_{\rm \cap L}=N_{\rm \cap
R}=0$), one has $C_{\rm i}+C_{\rm ii}+C_{\rm iii}+C_{\rm iv}=0$,
and the UCFs have the same magnitude as those for an asymmetric
dot.

To express ${\rm var}[g]$ for arbitrary $N_{\rm L}$, $N_{\rm R}$,
$N_{\rm \cap L}$, and $N_{\rm \cap R}$, we find it beneficial to
introduce the quantities $n_\kappa = N_\kappa/(N_{\rm L}+N_{\rm
R})$ and $w_\kappa = 1-N_{\cap\kappa}/N_\kappa$, where $\kappa
={\rm L,R}$. Making use of the fact that $n_{\rm L}+n_{\rm R}=1$,
we find
\begin{eqnarray}
\label{Eq:ucf-updown1}
{\rm var}[g] &=& {\rm var}[g]_{\rm asym} 
\\
&+& {2\over \beta}\, n_{\rm L}^2 n_{\rm R}^2
\left( {1-(1-n_{\rm L})w_{\rm L} - (1-n_{\rm R})w_{\rm R}\over
1+n_{\rm L}w_{\rm L} +n_{\rm R}w_{\rm R}} \right)^2
\nonumber
\end{eqnarray}
where in this notation ${\rm var}[g]_{\rm asym} = (2/\beta) n_{\rm
L}^2 n_{\rm R}^2$. In the special case where $N_{\rm L}= N_{\rm
R}$, displacing the leads suppresses the symmetry-induced
contribution to UCFs by a factor $\big[(2-w_{\rm L}-w_{\rm
R})/(2+w_{\rm L}+w_{\rm R})\big]^2$.

In terms of the original quantities $N_{\rm L}$, $N_{\rm R}$,
$N_{\rm \cap L}$, and $N_{\rm \cap R}$, Eq.~(\ref{Eq:ucf-updown1})
takes the form
\begin{eqnarray}
\label{Eq:ucf-updown2}
{\rm var}[g] &=& {\rm var}[g]_{\rm asym} 
\\
&+&
{2\over \beta}\,  {N_{\rm L}^2 N_{\rm R}^2 \over
( N_{\rm L}+ N_{\rm R})^4}
\left(
N_{\rm R}N_{\rm \cap L}/N_{\rm L} +   N_{\rm L}N_{\rm \cap R}/N_{\rm R}
\over 2N_{\rm L}+2N_{\rm R}-N_{\rm \cap L} -N_{\rm \cap R} \right)^2
\nonumber 
\end{eqnarray}
where ${\rm var}[g]_{\rm asym} = (2/\beta) N_{\rm L}^2 N_{\rm R}^2
( N_{\rm L}+ N_{\rm R})^{-4}$. Comparison with Eq.~(\ref{Eq:g_UD})
shows that lead displacement suppresses the symmetry-induced
contributions to UCFs by a factor that is the square of the
suppression of the symmetry-induced contributions to the average
conductance.

\subsection{UCFs in a left-right or inversion-symmetric quantum dot}

For a systems with a left-right or an inversion symmetry, we once
again find the magnitude of the UCFs by evaluating ${\rm
covar}[R,R']$.  For these symmetries, we must consider the
contributions in Fig.~\ref{Fig:ucf-asymleads2} in {\it addition}
to those in Fig.~\ref{Fig:ucf-asymleads1}. The origin of the extra
contributions in Fig.~\ref{Fig:ucf-asymleads2} is most clearly
understood by considering the case of perfectly symmetric leads.
Then the left-right and inversion symmetries map lead L onto lead
R, which means that if path 2 is paired with path 1' then path 2
will hit lead R when path 1' hits lead L (meaning the image of
path 1' hits lead R). One can thereby immediately see that the
contribution $C_{\rm v}$ in Fig.~\ref{Fig:ucf-asymleads2}
contributes to ${\rm covar}[R_{\rm L},R_{\rm R}]$ (this was not
the case for up-down symmetry, since there path 1' hits the same
lead as the image of path 1'). For asymmetric leads a similar
situation occurs. If path 1' hits the intersection region of width
$W_\cap$ on
 lead L then its image hits lead R; thus path
2 will  also hit lead R if it is paired with 1' over this segment.

The rules to evaluate each contribution are the same as for up-down
symmetry, with now necessarily $N_{\rm \cap L}=N_{\rm \cap
R}=N_\cap$.
Using these rules, we find that
\begin{eqnarray}
C_{\rm v}+C_{\rm vi} &=&
{2\over \beta}\,{4N_\cap N_{\rm L}N_{\rm R} - N_\cap^2(N_{\rm L}+N_{\rm R})
\over  (2N_{\rm L}+N_{\rm R}-2N_\cap)^2(N_{\rm L}+N_{\rm R})}, \quad \quad\\
C_{\rm vii}+C_{\rm viii} &=& - {2\over \beta}\,{2N_\cap N_{\rm
L}N_{\rm R} \over (2N_{\rm L}+N_{\rm R}-2N_\cap)(N_{\rm L}+N_{\rm
R})^2}.\quad \quad
\end{eqnarray}
Summing these contributions and writing the result with the same
denominator as Eq.~(\ref{Eq:ucf-updown2}) gives
\begin{eqnarray}
& & \hskip -10mm
C_{\rm v}+C_{\rm vi} + C_{\rm vii}+C_{\rm viii} 
\nonumber \\
&=& - {2\over
\beta}\,{N_\cap^2 (N_{\rm L}^2-N_{\rm R}^2)^2 
\over (2N_{\rm L}+N_{\rm R}-2N_\cap)^2(N_{\rm L}+N_{\rm R})^4}.
\label{Eq:extra-contributions-for-LR}
\end{eqnarray}
Adding this set of contribution to those already calculated in the
previous section, we find that the UCFs of a left-right or
inversion-symmetric dot with asymmetric leads is given by
\begin{eqnarray}
{\rm var}[g] &=& {\rm var}[g]_{\rm asym} 
\nonumber \\
&+& {2\over \beta}\, {N_{\rm
L}^2N_{\rm R}^2 \over (N_{\rm L}+N_{\rm R})^4} \left( {N_\cap
\over N_{\rm L}+N_{\rm R}-N_\cap}\right)^2
\label{Eq:ucf-leftright}.
\end{eqnarray}
By comparing this with Eq.~(\ref{Eq:g-LR-final-result}),
we find that the
suppression of symmetry-induced contributions to UCFs is the
square of suppression of the symmetry-induced contributions to
the average conductance (just as we already
found for an up-down symmetric system).

\subsection{UCFs in a 4-fold symmetric}

For completeness, we now briefly discuss UCFs in a 4-fold
symmetric dot with asymmetric leads. A 4-fold dot has all three of
the symmetries discussed above. Thus the UCFs in a four-fold
symmetric system are given by the sum of all possible
symmetry-induced contributions (just as with symmetric leads
\cite{WSK1}). Given the results in the preceding sections, the
general formula is easily determined. Here we give the result for
the special case $N_{\rm L}=N_{\rm R}=N$,
% \begin{widetext}
\begin{eqnarray}
{\rm var}[g] &=&
{1\over 8\beta}\Bigg[
  \left({N_{\rm \cap LR}\over 2N - N_{\rm \cap LR}}\right)^2
+ \left({N_{\rm \cap inv}\over 2N - N_{\rm \cap inv}}\right)^2
\nonumber \\
& & +\left({N_{\rm \cap UD:L}+N_{\rm \cap UD:R}
         \over 4N - N_{\rm \cap UD:L}-N_{\rm \cap UD:R}} \right)^2
+ 1  \Bigg]
\end{eqnarray}
% \end{widetext}
where $N_{\rm \cap LR}$ is the intersection between leads  L and R
under the left-right symmetry, $N_{\rm \cap inv}$ is the
intersection between leads L and R under the inversion symmetry,
and $N_{\rm \cap UD:L}$ ($N_{\rm \cap UD:R}$) is the intersection
of lead L (R)  with itself under the up-down symmetry. The final
term in the square-bracket represents the usual UCFs for an
asymmetric dot.

Note that the suppression of each symmetry-induced term goes like
the square of the equivalent term in the average conductance,
Eq.~(\ref{Eq:delta-g-4fold}).
%However, they appear with different
%signs in  Eq.~(\ref{Eq:delta-g-4fold}).

%============================
\section{\label{sect:RMT}Comparison to random-matrix theory}

In this section we compare the semiclassical predictions derived
in the previous sections to numerical results obtained from a
phenomenological random-matrix model. This model generalizes the
construction discussed in Section 9 of  part I (Ref.\
\cite{WSK1}).

The general framework is the same as in part I: The conductance is
obtained from the Landauer formula $g=\tr [t^\dagger
t]$, where $t$ is the transmission block of a scattering matrix $S=\left(%
\begin{array}{cc}
  r & t' \\
  t & r' \\
\end{array}%
\right)$ given by
\begin{equation}
S=P^T (1-FQ)^{-1} F P . \label{eq:dyns}
\end{equation}
Here, $F$ is an internal unitary evolution operator of dimension
$M$ while $P$ is an $M\times 2N$ dimensional matrix specified
below, and $Q=1-PP^T$.

In part I we assumed that the leads respect the geometrical
symmetries, which allows to fully desymmetrize the system. One
can then introduce a fixed form of the matrix $P$ and attribute
the effects of symmetries solely to the internal dynamics (the
resulting RMT ensembles for $F$ are given in Table 2 of Ref.\
\cite{WSK1}). It is clear that this full desymmetrization fails
when leads are to be displaced. For up-down symmetry, for
instances, desymmetrization identifies two effectively separate
systems (consisting of modes of even and odd parity) which do not
couple to each other. Shifting lead modes in this representation
has no effect since RMT is invariant under the permutation of
matrix indices. A real displacement of leads, however, mixes the
states of even and odd parity. The reason for this discrepancy is
that leads are defined locally in real space, while parity is a
global symmetry which connects remote parts of the system.

%%%%%%%%%%%%%%%%%%%%%%%%%%%%%%%%%%%%%%%%%%%%%%%%%%%%%%%%%%%%%%%%%%%%%%%%%%%%%
\begin{figure}
\hbox{\includegraphics[width=\columnwidth]{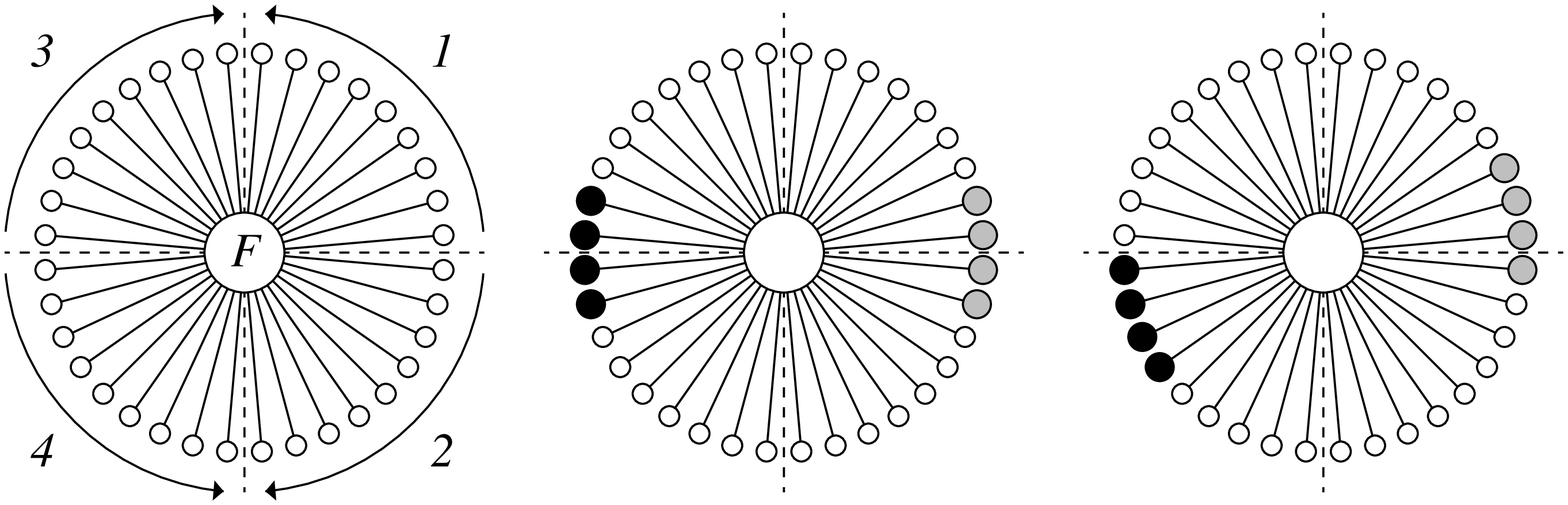}}
\caption[]{\label{fig:portmodel} 
Left panel: Model of a scatterer
(central circle) with internal evolution operator $F$, coupled to
ports to which modes of the leads can be attached. The labels
identify four segments, in which the ports are numerated in the
direction of the arrow (port $1$ to $M/4$ in segment 1, port
$M/4+1$ to $M/2$ in segment 2, port $M/2+1$ to $3M/4$ in segment
3, and port $3M/4+1$ to $M$ in segment 4). The dashed lines
indicate the possible lines of reflection symmetry. Middle and
right panels: Filled circles indicate ports coupled to the  left
lead, shaded circles indicate ports coupled to the  right lead.
Shown are a fully symmetry-respecting arrangement and an
arrangement in which both leads are displaced, respectively.
 }
\end{figure}
%%%%%%%%%%%%%%%%%%%%%%%%%%%%%%%%%%%%%%%%%%%%%%%%%%%%%%%%%%%%%%%%%%%%%%%%%%%%%

%%%%%%%%%%%%%%%%%%%%%%%%%%%%%%%%%%%%%%%%%%%%%%%%%%%%%%%%%%%%%%%%%%%%%%%%%%%%%
\begin{table*}
\begin{tabular}{|l|c c|}
\hline \hline & $ B=0 $& $B\gg B_c$\\  \hline
no spatial sym. &     COE($M$)        &               CUE($M$)\\
left-right  sym. & $A^\dagger$ COE$^2$($M/2$) $A$ & $A^\dagger$ COE($M$) $A$ \\
inversion sym.   &    $D A^\dagger$ COE$^2$($M/2$) $A D$ & $D
A^\dagger$ CUE$^2$($M/2$) $A D$
\\
up-down  sym.& $CA^\dagger$ COE$^2$($M/2$) $AC$ & $CA^\dagger$ COE($M$) $AC$\\
four-fold  sym.
& $\ D A^\dagger$[$A^\dagger$ COE$^2$($M/4$) $A$]$^2$$A D \ \ $ 
& $\ \ D A^\dagger$ [$A^\dagger$ COE($M/2$) $A$]$^2$$A D \ $
\\
\hline\hline
\multicolumn{3}{|r|}{$\phantom{\begin{array}{c} 1 \\ 1 \\ 1 \\ 1 \\ 1\end{array}}$
with 
$A=2^{-1/2} \left(\begin{array}{cc} 1 & 1 \\  i & -i \\ \end{array} \right)$,
 $C=\left(\begin{array}{cccc} 1&0&0&0 \\ 0&0&1&0 \\ 0&1&0&0 \\
0&0&0&1 \end{array} \right)$
and 
$D=\left(\begin{array}{cccc} 1&0&0&0 \\ 0&1&0&0 \\ 0&0&0&1 \\
0&0&1&0 \end{array} \right)$
}
\\
\hline\hline
\end{tabular}
\caption{
Random-matrix ensembles for the internal evolution
operator $F$ in a basis which is suitable for displacing the leads
(see Fig.\ \ref{fig:portmodel}). The different entries refer to
 various geometric symmetries in absence or presence of a
magnetic field. We only consider the case $M \mbox{ mod 4}=0$.
Block composition of two identical matrix ensembles of dimension
$M$ is abbreviated as $X^2(M)=X(M)\otimes X(M)$.
%; in the general case one encounters composition of ensembles with dimensions that differ at most by 1.
} \label{tab:f}\end{table*}
%%%%%%%%%%%%%%%%%%%%%%%%%%%%%%%%%%%%%%%%%%%%%%%%%%%%%%%%%%%%%%%%%%%%%%%%%%%%%

It is therefore necessary to define both the internal evolution
operator $F$ as well as the coupling to the leads $P$ in a way
which resembles modes in a real-space basis. In principle, this
can be done, e.g., based on the sinusoidal transverse mode
profiles of a strip resonator. We adopt a similar, but more
efficient procedure, whose principle idea is shown in Fig.\
\ref{fig:portmodel}. The illustration shows an abstract scatterer
with $M$ ports which serve as possible contacts to the system. For
each lead we select $N$ ports (with index $i_n$ for lead L and
$j_n$ for lead R); the remaining ports are closed off. The
internal evolution operator $F$ describes the transport from port
to port. The scattering matrix is then given by Eq.\
(\ref{eq:dyns}) where $P_{mn}=\delta_{m,i_n}+\delta_{m,j_{n-N}}$.

%%%%%%%%%%%%%%%%%%%%%%%%%%%%%%%%%%%%%%%%%%%%%%%%%%%%%%%%%%%%%%%%%%%%%%%%%%%%%
\begin{figure*}
\centerline{\hbox{\includegraphics[width=0.8\textwidth]{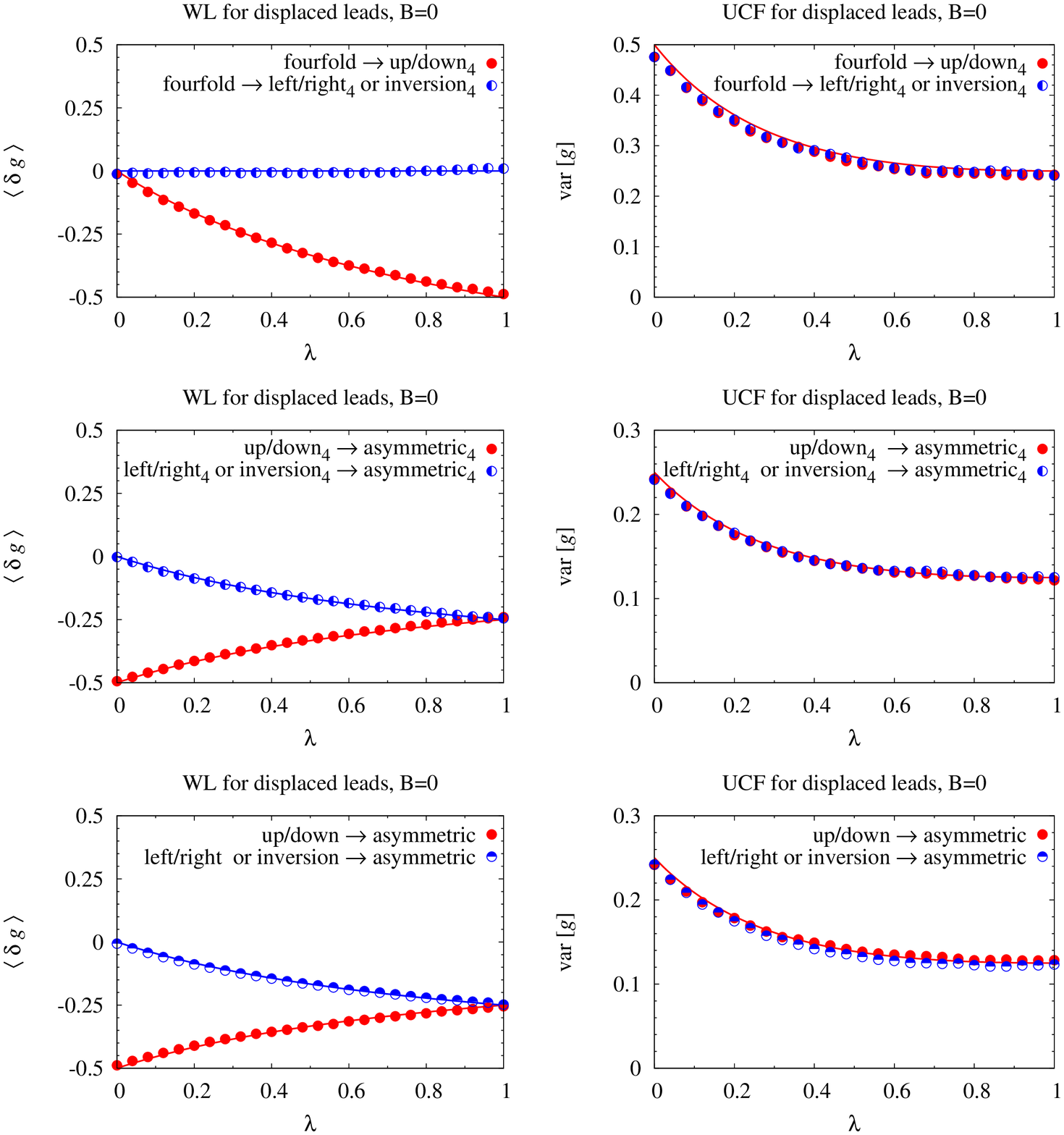}}}
% {henning-numerics/floquet_offset_rmt_50_1000_b0.eps}}}
\caption[]{\label{Fig:numerics1a} 
(colour online). Weak localization correction
(WL, left panels) and  universal conductance fluctuations (UCF,
right panels) as a function of the displacement of both leads from
their symmetry-respecting positions for systems with fixed
internal symmetry. The displacement is measured in terms of
$\lambda=1-W_\cap/W$. The data points (circles with a variety of
filling styles) are obtained from an average over 5000
realizations of the RMT model described in the text ($M=1000$,
$N=50$). The curves show the semiclassical prediction
(\ref{eq:dglambda}) for WL and (\ref{eq:varglambda}) for UCF.
Labels `$A\to B$' specify the symmetry of the lead arrangement at
$\lambda=0$ (symmetric arrangement) and $\lambda=1$ (where at
least one of the symmetries is fully removed). In these labels,
the subscript $_4$ on $A$ or $B$ indicates that the internal
symmetry is four-fold; if this subscript is not present the
internal symmetry is identical to the one specified by $A$. In
this figure, the magnetic field is set to $B=0$.
 }
\end{figure*}
%%%%%%%%%%%%%%%%%%%%%%%%%%%%%%%%%%%%%%%%%%%%%%%%%%%%%%%%%%%%%%%%%%%%%%%%%%%%%

%%%%%%%%%%%%%%%%%%%%%%%%%%%%%%%%%%%%%%%%%%%%%%%%%%%%%%%%%%%%%%%%%%%%%%%%%%%%%
\begin{figure*}
\centerline{\hbox{\includegraphics[width=0.8\textwidth]{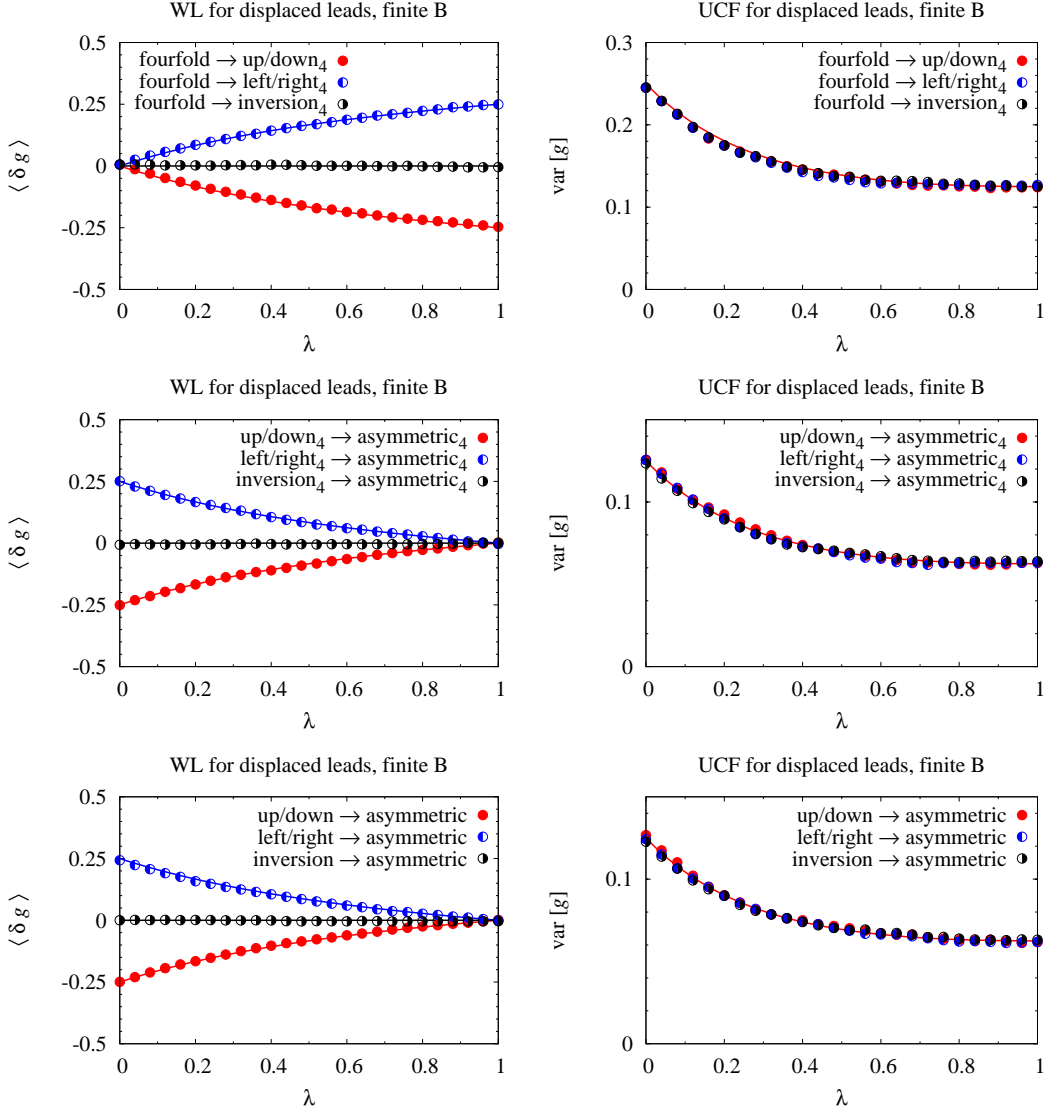}}}
% {henning-numerics/floquet_offset_rmt_50_1000_bfinite.eps}}}
\caption[]{\label{Fig:numerics1b} 
(colour online). Same as Fig.\
\ref{Fig:numerics1a}, but for a finite magnetic field.
 }
\end{figure*}
%%%%%%%%%%%%%%%%%%%%%%%%%%%%%%%%%%%%%%%%%%%%%%%%%%%%%%%%%%%%%%%%%%%%%%%%%%%%%

%%%%%%%%%%%%%%%%%%%%%%%%%%%%%%%%%%%%%%%%%%%%%%%%%%%%%%%%%%%%%%%%%%%%%%%%%%%%%
\begin{figure*}
\centerline{\hbox{\includegraphics[width=0.8\textwidth]{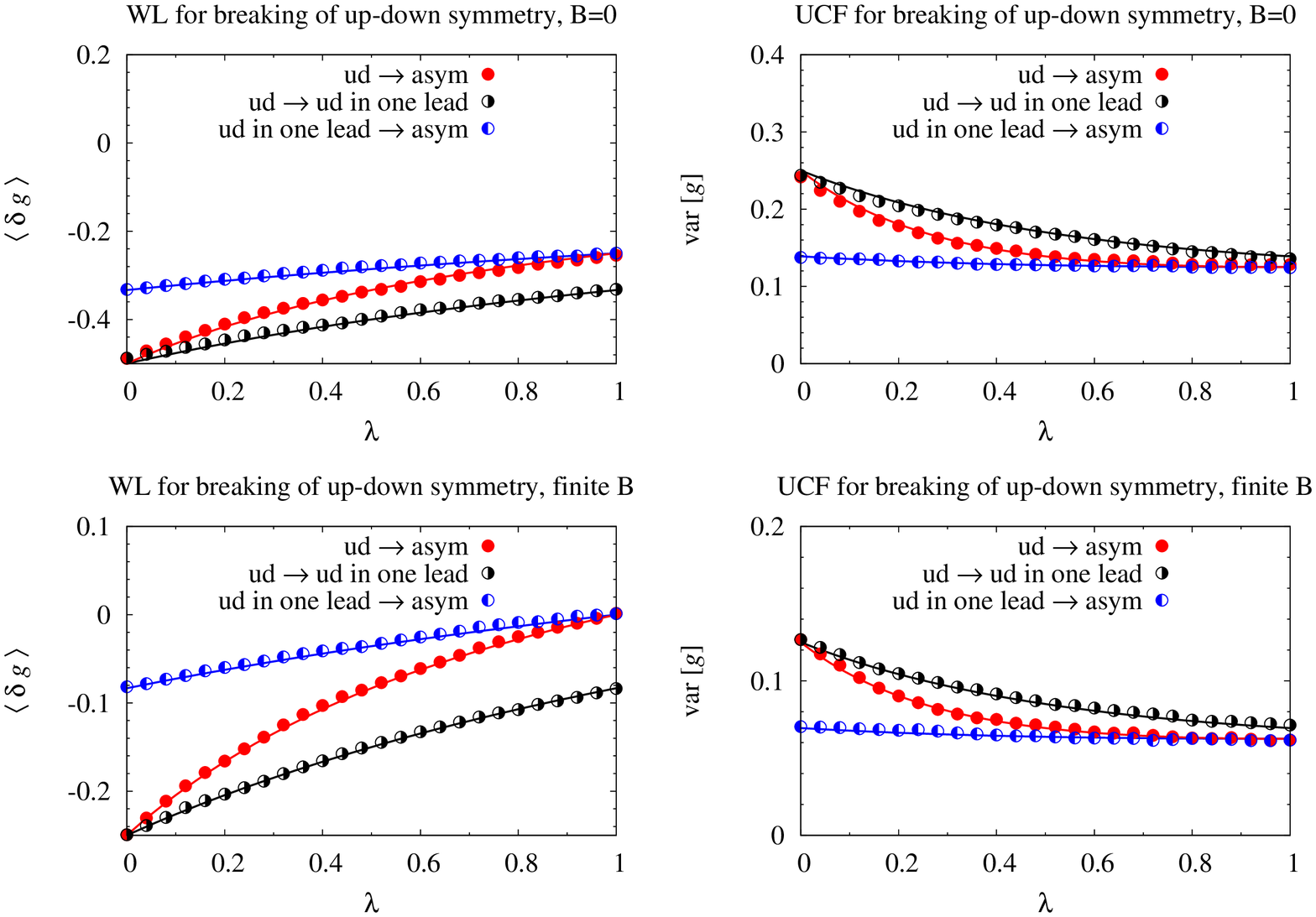}}}
% {henning-numerics/floquet_offset_rmt_50_1000_all_updown.eps}}}
\caption[]{\label{Fig:numerics2} 
(colour online). Same as Figs.\
\ref{Fig:numerics1a} and \ref{Fig:numerics1b}, but comparing the
displacement of both leads for internal up-down symmetry (solid
circles) to the displacement of the first lead (circles filled on
the right), followed by the displacement of the second lead
(circles filled on the left).
 }
\end{figure*}
%%%%%%%%%%%%%%%%%%%%%%%%%%%%%%%%%%%%%%%%%%%%%%%%%%%%%%%%%%%%%%%%%%%%%%%%%%%%%

A crucial point of the illustration in Fig.\ \ref{fig:portmodel}
is the numeration of ports, which are grouped into 4 segments that
map in specific ways onto each other when symmetry operations are
applied. (i) Left-right symmetry maps segment 1 onto segment 3 and
segment 2 onto segment 4.
 (ii)
Up-down symmetry maps segment 1 onto segment 2 and segment 3 onto
segment 4. (iii) Inversion symmetry maps segment 1 onto segment 4
and segment 2 onto segment 3. (iv) Four-fold symmetry maps all
segments onto each other.

In the basis of these ports, the explicit symmetries of the
internal evolution operator $F$ are specified in Table
\ref{tab:f}. Since up-down and left-right symmetry are both
manifestations of a reflection symmetry, they are now simply
related by a interchanging segments 2 and 3 (as described by the
matrix $C$ defined in Table \ref{tab:f}); this is a consequence of
the fact that we do not fully desymmetrize the up-down symmetry
(the left-right symmetric case can never be fully desymmetrized
because one has to keep track of the identity of the leads). A
finite magnetic field breaks these symmetries, but still allows
one to define a generalized time-reversal symmetry. Similarly, for
vanishing magnetic field, inversion symmetry is obtained from
reflection symmetry by interchanging segments 3 and 4 (as
described by the matrix $D$ defined in the table caption). The
slightly different systematics in the presence of a magnetic field
arises because the orientation of the segments matters;
consequently, for inversion symmetry, time-reversal symmetry is
effectively broken but the geometric symmetry itself is still
present in the dynamics (trajectories still occur in
symmetry-related pairs).

A convenient choice of a fully symmetry-respecting arrangement of
leads which applies to all internal symmetries is given by
\begin{equation}
P=\left(%
\begin{array}{cccc}
  1_{{\cal N}\times {\cal N}} & 0_{{\cal N}\times {\cal N}} & 0_{{\cal N}\times {\cal N}} &  0_{{\cal N}\times {\cal N}} \\
  0_{{\cal M}\times {\cal N}} & 0_{{\cal M}\times {\cal N}} & 0_{{\cal M}\times {\cal N}}& 0_{{\cal M}\times {\cal N}} \\
  0_{{\cal N}\times {\cal N}} & 1_{{\cal N}\times {\cal N}} & 0_{{\cal N}\times {\cal N}} &  0_{{\cal N}\times {\cal N}} \\
  0_{{\cal M}\times {\cal N}} & 0_{{\cal M}\times {\cal N}} & 0_{{\cal M}\times {\cal N}}& 0_{{\cal M}\times {\cal N}} \\
  0_{{\cal N}\times {\cal N}} & 0_{{\cal N}\times {\cal N}} & 1_{{\cal N}\times {\cal N}} &  0_{{\cal N}\times {\cal N}} \\
  0_{{\cal M}\times {\cal N}} & 0_{{\cal M}\times {\cal N}} & 0_{{\cal M}\times {\cal N}}& 0_{{\cal M}\times {\cal N}} \\
  0_{{\cal N}\times {\cal N}} & 0_{{\cal N}\times {\cal N}} & 0_{{\cal N}\times {\cal N}} & 1_{{\cal N}\times {\cal N}} \\
  0_{{\cal M}\times {\cal N}} & 0_{{\cal M}\times {\cal N}} & 0_{{\cal M}\times {\cal N}}& 0_{{\cal M}\times {\cal N}} \\
\end{array}%
\right),
\end{equation}
where ${\cal N}=N/2$ and ${\cal M}=M/4-N/2$.
 The case of a four-fold symmetry in principle allows two
symmetry-respecting arrangements (aligned along each of the two
symmetry lines of reflection); these two arrangements are
equivalent in RMT and again related by a reshuffling of the 4
segments. The form of $P$ for generally placed leads is easily
read off Fig.\ \ref{fig:portmodel}.

Figures \ref{Fig:numerics1a} (for $B\ll B_{\rm c}$) and
\ref{Fig:numerics1b} (for $B\gg B_{\rm c}$) show  how the
weak localization correction and universal conductance
fluctuations are affected when the leads are moved away from the
symmetry-respecting positions. The degree of displacement is
quantified by a variable $\lambda=1-W_\cap/W$ ($\lambda=0$ in the
symmetric arrangement, $\lambda=1$ in the asymmetric arrangement).
The data points are based on an ensemble average over $5000$ RMT
matrices with $M=1000$ and $N=50$, while the curves are the
predictions of our semiclassical theory, which can be written as
\begin{equation}
\delta g(\lambda) = \delta g(1) +[\delta g(0)-\delta
g(1)]\frac{1-\lambda}{1+\lambda}, \label{eq:dglambda}
\end{equation}
\begin{equation}
{\rm var} g(\lambda) ={\rm var} g(1) +[{\rm var} g(0)-{\rm var}
g(1)]\left(\frac{1-\lambda}{1+\lambda}\right)^2.
\label{eq:varglambda}
\end{equation}
Starting from a four-fold symmetry, leads can be displaced in a
manner which still preserves left-right, inversion, or up-down
symmetry. To preserve up-down symmetry alone, 
one can imagine splitting one lead in two
and moving the two parts in opposite directions (both parts would
remain contacted to the same source or drain electrode).
The remaining symmetry of the lead
arrangement can then be broken  by further displacement of the
leads. In the figures, the subscript $_4$ is used to distinguish
these  situations (in which the underlying internal symmetry is
four-fold) from the symmetry breaking in systems with only a
single internal symmetry. E.g., the label `left/right$_4\to$
asymmetric$_4$' refers to the displacement of leads out of a
left-right symmetric position where the internal symmetry is
four-fold, while the label `left/right $\to$ asymmetric' refers to
the displacement of leads out of a left-right symmetric position
where the internal symmetry is itself only left-right symmetric.
According to our theory, the weak localization correction should
behave identically in both situations; this also applies to the
UCFs. This statement is validated by the numerical data. Indeed,
excellent agreement of the numerical data with the semiclassical
predictions is observed in all cases.

As discussed earlier in this paper,  in the up-down symmetric case
it is interesting to displace only one lead while the other lead
remains on the symmetry line (the symmetry-preserving positions in
the up-down symmetric case are absolute, in contrast to the
left-right symmetric case where these positions are relative to
each other). The effect on the transport is shown in Fig.\
\ref{Fig:numerics2}, along with the effect of the consecutive
displacement of the second lead, and the simultaneous displacement
of both leads. According to our theory, the effects of consecutive
displacement of the leads are cumulative: The displacement of the
first lead is described by Eqs.\ (\ref{eq:dglambda}),
(\ref{eq:varglambda}) with $\lambda\to\lambda/2$ (covering the
range [0,1/2]), while the displacement of the second lead
completes the transition according to the substitution
$\lambda\to(1+\lambda)/2$ (covering the range [1/2,1]). The
numerical results are in perfect agreement with this prediction.

We conclude with some additional remarks on the RMT model. For
leads which respect the symmetries, the construction presented
here is equivalent to the model presented in part I (which then is
more efficient); this equivalence also extends to the symmetry
breaking in the internal dynamics, which then requires to
interpolate between ensembles of Table \ref{tab:f}. Following
earlier works, the RMT model can be further utilized to include
the effects of dephasing and a finite Ehrenfest time. For
dephasing, this is achieved by opening additional ports which
couple to a voltage probe \cite{buttiker} or a dephasing stub
\cite{Brouwer-Frahm-Beenakker}. A finite Ehrenfest time is
obtained when $F$ represents a dynamical system, such as the
kicked rotator \cite{Schomerus-Jacquod} (which also possesses
discrete symmetries). This strategy can also be used to probe the
case of dynamics which are not fully chaotic (which in the kicked
rotator is achieved for moderate values of the kicking strength).

%============================
\section{Concluding remarks}

\label{sect:conclusions}

The transport calculations performed here assume that the
classical dynamics is uniformly chaotic, and in particular do not
apply to system with islands of stability in phase space (such as
the annular billiard studied in Refs.\
\cite{Schomerus-Beenakker,Sim-Schomerus}), or networks of chaotic
dots inter-connected by narrow leads (such as the double dot in
Ref.~\cite{WMM}). It would be intriguing to study the shape of the
back- and forward-scattering peaks for such systems.
%Perhaps the
%peak-shape would be sensitive to details of the dynamics that are
%difficult to access by other transport properties (such as the
%average conductance, etc).

\section{Acknowledgements}
RW thanks P.~Brouwer, P.~Marconcini and M.~Macucci for interesting 
and useful discussions.  RW and HS are grateful for the hospitality of the
Banff International Research Station, where this work was initiated.

%%%%%%%%%%%%%%%%%%%%%%%%%%%%%%%%%%%%%%%%%%%%%%%%%%%
\appendix
\section{Obtaining UCFs from the variance of reflection}
\label{appendix}

In Section~\ref{sect:UCFs} we pointed out that unitarity implies
${\rm covar}[R_{\rm L},R_{\rm R}]= {\rm var}[R_{\rm L}]={\rm
var}[R_{\rm R}]$. Here we outline a semiclassical calculation of
${\rm var}[R_{\rm L}]$, which acts as a check of the semiclassical
calculation of ${\rm covar}[R_{\rm L},R_{\rm R}]$ in
Section~\ref{sect:UCFs}. The  rules to calculate each contribution
remain the same as for ${\rm covar}[R_{\rm L},R_{\rm R}]$.
However, the contributions that we consider differ by the
requirement that all paths start and end on the same lead L.

We know that the result must be invariant under the interchange of
labels ``L'' and ``R'', and this invariance is manifestly obvious
in the contributions to ${\rm covar}[R_{\rm L},R_{\rm R}]$. In
contrast, this invariance is  hidden in the contributions to 
${\rm var}[R_{\rm L}]$ that we discuss here. 
Thus the simplest check that one has not missed any
contributions is that this invariance is present
when one sums the contributions.

\subsection{Up-down symmetric dot}
In the case of an up-down symmetric dot, all contributions in both
Fig.~\ref{Fig:ucf-asymleads1} and Fig.~\ref{Fig:ucf-asymleads2}
contribute to ${\rm var}[R_{\rm L}]$ once we change all lead
labels so that ``R'' $\to$ ``L'' and ``$\cap$R'' $\to$ ``$\cap$L''
(but not vice versa). Writing contributions to ${\rm var}[R_{\rm
L}]$ with a ''prime''(to distinguish them from contributions to
${\rm covar}[R_{\rm L},R_{\rm R}]$) we find
\begin{widetext}
\begin{eqnarray}
C'_{\rm v} + C'_{\rm vi} + C'_{\rm i} 
&=& {2 \over \beta}\, {N_{\rm \cap L}^2(N_{\rm L}+N_{\rm
R})^2 + 4N_{\rm \cap L}(N_{\rm L}-N_{\rm \cap L})N_{\rm L}(N_{\rm
L}+N_{\rm R}) + 4(N_{\rm L}-N_{\rm \cap L})^2N_{\rm L}^2 \over
(2N_{\rm L}+2N_{\rm R}-N_{\rm \cap L}-N_{\rm \cap R})^2 (N_{\rm
L}+N_{\rm R})^2}, 
\\
C'_{\rm vii} + C'_{\rm viii} + C'_{\rm ii}+C'_{\rm iii} 
&=& -{2 \over \beta}\, {2N_{\rm L}^2 \big[N_{\rm \cap
L}(N_{\rm L}+N_{\rm R}) + 2 (N_{\rm L}-N_{\rm \cap L})N_{\rm
L}\big] \over (2N_{\rm L}+2N_{\rm R}-N_{\rm \cap L}-N_{\rm \cap
R}) (N_{\rm L}+N_{\rm R})^3},
\\
C'_{\rm iv} &=& {2 \over \beta}\, {N_{\rm L}^4 \over (N_{\rm
L}+N_{\rm R})^4}.
\end{eqnarray}
\end{widetext}

As in section~\ref{sect:UCFs}, we find that this sum is most
easily evaluated by re-writing the contributions in terms
$n_\kappa = N_\kappa/(N_{\rm L}+N_{\rm R})$ and $w_\kappa =
1-N_{\cap\kappa}/N_\kappa$ for $\kappa ={\rm L,R}$. Performing a
little algebra using $n_{\rm L}+n_{\rm R}=1$, we then recover
Eq.~(\ref{Eq:ucf-updown1}), and therefore ${\rm var}[R_{\rm
L}]={\rm covar}[R_{\rm L},R_{\rm R}]$. Furthermore, expression
Eq.~(\ref{Eq:ucf-updown1})
 is invariant under the
interchange of labels ``L'' and ``R'', which entails  ${\rm
var}[R_{\rm R}]={\rm var}[R_{\rm L}]$. Thus the semiclassical
method obeys the relations ${\rm var}[R_{\rm L}]={\rm var}[R_{\rm
R}] ={\rm covar}[R_{\rm L},R_{\rm R}]$, as required by the
unitarity of the scattering matrix. This strongly suggests that we
have not missed any contributions and gives us confidence in the
result; particularly, it is noteworthy that the individual
contributions in ${\rm var}[R_{\rm L}]$ and ${\rm covar}[R_{\rm
L},R_{\rm R}]$ combine in very different ways to give the
invariance under the interchange of ``L'' and ``R''.

\subsection{Left-right or inversion-symmetric dot}

The evaluation of ${\rm var}[R_{\rm L}]$ for a left-right or
inversion-symmetric dot is very similar to that for an up-down
symmetric dot.  However, here, when a path hits the L lead then
its image hits the R lead.  This means that there are no
contributions to ${\rm var}[R_{\rm L}]$ of the form shown in
Fig.~\ref{Fig:ucf-asymleads2}, since all paths must go from the L
lead to the L lead. Thus to get ${\rm var}[R_{\rm L}]$ for a
left-right or inversion-symmetric dot, we need to subtract those
contributions from the result for ${\rm var}[R_{\rm L}]$ in an
up-down symmetric dot. The sum of these contributions to ${\rm
var}[R_{\rm L}]$, written with the same denominator as in
Eq.~(\ref{Eq:ucf-updown2}), is
\begin{eqnarray}
& & \hskip -10mm
C'_{\rm v}+C'_{\rm vi} + C'_{\rm vii}+C'_{\rm viii} 
\nonumber \\
&=&
{2\over \beta}\,{N_\cap^2 (N_{\rm L}^2-N_{\rm R}^2)^2 \over
(2N_{\rm L}+N_{\rm R}-2N_\cap)^2(N_{\rm L}+N_{\rm R})^4}.
\end{eqnarray}
This only differs by an overall sign from the sum of contributions
in Eq.~(\ref{Eq:extra-contributions-for-LR}). Subtracting this
from the result Eq.~(\ref{Eq:ucf-updown2}), we get ${\rm
var}[R_{\rm L}]$ for a left-right or inversion-symmetric dot. The
result equals ${\rm covar}[R_{\rm L},R_{\rm R}]$ given by
Eq.~(\ref{Eq:ucf-leftright}), thus we have ${\rm covar}[R_{\rm
L},R_{\rm R}]={\rm var}[R_{\rm L}]={\rm var}[R_{\rm R}]$, as
required by the unitarity of the scattering matrix.

%=======================================================================

%%%%%%%%%%%%%%%%%%%%%%%%%%%%%%%%%%%%%%%%%%%%%%%%%%%%%%%%%%%%%%%%%%%%%%
%%%%%%%%%%%%%%%%%%%%%%%%%%%%%%%%%%%%%%%%%%%%%%%%%%%%%%%%%%%%%%%%%%%%%%
%%%%%%%%%%%%%%%%%%%%%%%%%%%%%%%%%%%%%%%%%%%%%%%%%%%%%%%%%%%%%%%%%%%%%%


\begin{thebibliography}{10}
\bibitem{footnote:failed-cbs}
The fact that the paths in contributions 2i-iv reflect back into
the dot slightly changes the action difference between them. It
was shown in the context of failed coherent backscattering that
this change of the action difference is irrelevant \cite{Whi07}.
\bibitem{footnote:tunnel-wl}
This is rather different from the weak localization correction
with tunnel barriers \cite{Whi07}.  There the survival probability
is only different near the encounter, which leads to an extra
prefactor of the ratio of dwell times (of the form $\tau_{\rm
D1}/\tau_{\rm D2}$). There is no such extra prefactor here.
\end{thebibliography}
\end{document}